\documentclass[nofootinbib,
amsmath,amssymb,
 aps,
prb,
twocolumn,
]{revtex4-2}

\usepackage{microtype}

\usepackage[colorlinks=true,citecolor=blue,linkcolor=magenta]{hyperref}
\usepackage{graphicx}
\usepackage[normalem]{ulem}
\usepackage{xcolor}
\ifdefined\added
\definechangesauthor[name={Geremia Massarelli}, color=BrickRed]{GM}
\definechangesauthor[name={Arijit Haldar}, color=NavyBlue]{AH}
\fi
\usepackage[mathlines]{lineno}

\def\HBBH {{\hat{H}^\textup{spin}_\textup{tBBH}}} \def\HBBHII {{\hat{H}^\textup{spin}_\textup{tSOC}}} \def\HJS {{\hat{H}^\textup{spin}_\textup{tSOC}}}

\def\opHtripI {{\hat{H}_\textup{tBBH}}}
\def\HtripII {\hat{H}_\textup{tSOC}}
\def\opHtripII {{\hat{H}_\textup{tSOC}}}
\def\Blkgap {{\Delta_\textup{bulk}}}
\def\modI {{tBBH}}
\def\modII {{tSOC}}

\def\HI {{\hat{H}_I}}
\def\HII {{\hat{H}_{I\!I}}}

\def\hs {{h_\text{s}}}
\def\hcstag {{h^{\text{(c)}}_\text{s}}}
\def\hchomI  {{h_\text{c1}}}
\def\hchomII {{h_\text{c2}}}

\newcommand{\mbf}[1]{\mathbf{#1}}
\newcommand{\fig}[1]{Fig.~\ref{#1}}

\newcommand{\figs}[1]{Figs.~\ref{#1}}

\newcommand{\Sec}[1]{Sec.~\ref{#1}}
\newcommand{\eqn}[1]{Eq.~\ref{#1}}

\newcommand{\eqns}[1]{Eqs.~\ref{#1}}

\newcommand{\transpose}{\top}

\definecolor{orange}{rgb}{1,0.5,0}

\definecolor{pink}{HTML}{f52c86}

\newcommand{\mycite}[1]{\cite{#1}}

\newcommand{\app}[1]{app.~\ref{#1}}

\def\useapp {}     

\def\pdg {{\vphantom\dagger}}
\def\dg {{\dagger}}
\def\ci {{\textup{i}}}
\def\re {{\mathrm{e}}}
\def\rd {{\mathrm{d}}}
\def\ua{\uparrow}
\def\da{\downarrow}
\def\singlet{(\ket{\!\uparrow\downarrow}-\ket{\!\downarrow\uparrow})/\sqrt{2}}
\def\td {t^\dagger}
\def\sd {s^\dagger}
\def\vac {|0\rangle}
\def\S {\mathbf{S}}
\newcommand{\Sc}[1]{S^{#1}}
\newcommand{\Scs}[2]{S^{(#1,#2)}}   \def\J {J_\mathrm{D}}
\def\Kv {J^y}
\def\Kh {J^x}
\def\Khv {J^{x,y}}
\def\K {J}
\def\Mx {{\mathcal{M}_x}}
\def\My {{\mathcal{M}_y}}
\def\R {{\mathbf{R}}}

\newcommand{\site}[2]{{#1,#2}}
\newcommand{\ket}[1]{{|#1\rangle}}
\newcommand{\bra}[1]{{\langle #1|}}
\def\Fid {\Phi^\dagger}
\def\Fi {\Phi^\pdg}
\def\Sid {\Psi^\dagger}
\def\Si {\Psi}
\def\TR {\mathcal{T}}
\def\M {\mathbb{M}}
\def \l {\ell}

\newcommand{\ignore}[1]{}

\def\k {{\mbf{k}}}

\def\dk {{\Delta \k}}
\def\W {{\mathcal{W}^\alpha_\k}}
\def\Fak {F_{\alpha\k}}

\def\WL {\mathcal{W}}
\def\WH {{H_{\WL^\alpha}}}
\newcommand{\WHc}[1]{{H_{\WL^{#1}}}}
\def\WB {{\nu^\alpha_j}}

\def\WBjs {{\nu^\alpha_{j=1,2}}}
\newcommand{\WBc}[1]{{\nu^{#1}_j}}
\newcommand{\WBcj}[2]{{\nu^{#1}_{j=#2}}}
\def\WS {{\ket{w^\alpha_j(\k)}}}

\newcommand{\bWSkc}[1] {{\bra{w^\alpha_j(#1)}}}

\newcommand{\WSck}[1]{\ket{w^{#1}_j(\k)}}
\def\nWl {{\tilde{\mathcal{W}}^{j}}}
\def\Fpol {{F^{\textup{pol}}}}

\def\p {{\mbf{p}}}

\def\pyj {{p^{(\nu^{x}_{j})}_y}}
\newcommand{\pyjb}[1]{p^{(\nu^{x}_{#1})}_y}

\ifdefined\added
\else
\newcommand{\added}[2][]{\textcolor{blue}{#2}\textsuperscript{\small\textcolor{red}{#1}}}
\definecolor{shadecolor}{RGB}{80,100,80}
\definecolor{pink}{RGB}{220,100,100}
\usepackage{marginnote}
\newcounter{mparcnt}

\fi

\newcommand{\br}{{\bf r}}
\newcommand{\bR}{{\bf R}}
\def\Hdim {\hat{H}_\mathrm{D}}

\def\spinor {vector}

\newcommand{\m}[1]{\mathcal{M}_{#1}}
\newcommand{\mx}{\m{x}}
\newcommand{\my}{\m{y}}
\newcommand{\cconj}{K}

\def\HLEI {\tau_3 H_\text{cont}^\text{\modI}}
\def\HLEII {\tau_3 H_\text{cont}^\text{\modII}}
\def\HLEIIeff {\tilde{H}_\text{cont}^\text{\modII}}

\def\HLEItop { \tau_3 H_\text{top}^\text{\modI} }
\def\HLEIright { \tau_3 H_\text{right}^\text{\modI} }

\def\HLEIIefftop { \tilde{H}_\text{top}^\text{\modII} }
\def\HLEIIeffright { \tilde{H}_\text{right}^\text{\modII} }

\newcommand{\psitop}{\psi_{\text{top}}}
\newcommand{\psiright}{\psi_{\text{right}}}
\newcommand{\psitilderight}{\tilde{\psi}_{\text{right}}}

\newcommand{\psitr}{\psi_{\text{top-right}}}
\newcommand{\psirt}{\psi_{\text{right-top}}}
\newcommand{\psitildert}{\tilde{\psi}_{\text{right-top}}}

\newcommand{\kgap}{\k_0}

\def\Hhs{H_{\hs}^\text{tSOC}(\k)}
\def\Hh{H_h^\text{tSOC}(\k)}

\def\gammatop{\gamma_\mathrm{top}}
\def\gammaright{\gamma_\mathrm{right}}
\def\gammatopright{\gamma_\mathrm{top-right}}
\def\gammarighttop{\gamma_\mathrm{right-top}}

\def\P{\mathcal{P}}

\newcommand{\Jtensor}{\mathbb{J}}
\newcommand{\KI}{{ \mathbb{J}_\text{intra}^{1,2} }}
\newcommand{\KII}{{ \mathbb{J}_\text{inter}^{2,1} }}
\newcommand{\C}{{ \mathbb{J}_\text{inter}^{1,1} }}
\newcommand{\D}{{ \mathbb{J}_\text{inter}^{2,2} }}

\newcommand{\Jintra}{\Jtensor_\text{intra}}
\newcommand{\Jinter}{\Jtensor_\text{inter}} \preprint{APS/123-QED}
\begin{document}
\def\mytitle{Higher-order topology and corner triplon excitations in two-dimensional quantum spin-dimer models}
\title{\mytitle}

\author{Arijit Haldar}
\email{arijit.haldar@utoronto.ca}
\affiliation{Department of Physics, University of Toronto, 60 St. George Street, Toronto, Ontario M5S 1A7 Canada}

\author{Geremia Massarelli}\email{gmassare@physics.utoronto.ca}
\affiliation{Department of Physics, University of Toronto, 60 St. George Street, Toronto, Ontario M5S 1A7 Canada}

\author{Arun Paramekanti}
\email{arun.paramekanti@utoronto.ca}
\affiliation{Department of Physics, University of Toronto, 60 St. George Street, Toronto, Ontario M5S 1A7 Canada}

\date{\today}

\begin{abstract}
The concept of free fermion topology has been generalized to $d$-dimensional phases that exhibit $(d-n)$-dimensional boundary modes, such as zero-dimensional (0D) corner excitations. Motivated by recent extensions of these ideas to magnetic systems, we consider 2D quantum paramagnets formed by interacting spin dimers with dispersive triplet excitations. We propose two examples of such dimer models, where the spin-gapped bosonic triplon excitations are shown to host bands with nontrivial higher-order topology.
We demonstrate this using real-space Bogoliubov--de Gennes calculations that reveal the existence of near-mid-bandgap corner triplon modes as a signature of higher-order bulk topology. We provide an understanding of the higher-order topology in these systems via a computation of bulk topological invariants as well as the construction of edge theories, and study their phase transitions as we tune parameters in the model Hamiltonians. We also discuss possible experimental approaches for detecting the emergent corner triplon modes.
\end{abstract}

\maketitle

\section{Introduction}

Higher-order topological (HOT) systems have emerged as the next step in understanding the rapidly growing family of topological phases in condensed matter physics. As a natural generalization of conventional or first-order topology, an $n$\textsuperscript{th}-order topological system in $d$-dimensions hosts protected gapless modes on any of its $(d-n)$-dimensional boundaries. 
For instance, a three-dimensional (3D) HOT system exhibiting second-order topology hosts protected hinge modes, while one exhibiting third-order topology hosts protected corner modes.
Similarly, a 2D HOT system showing second-order topology hosts localized corner modes. First discovered in the context of electric-multipolar insulators~\mycite{benalcazar2017quantized} in fermionic systems with quantized moments, the list of HOT systems was quickly expanded to include chiral and helical HOT insulators~\mycite{SchindlerHOT2018}. The core idea behind fermionic HOT insulators has recently been generalized, either theoretically or experimentally, to alternate platforms such as mechanical~\mycite{serra2018nature}, electrical~\mycite{imhof2018topolectrical}, microwave~\mycite{peterson2018quantized}, photonic~\mycite{el2019light}, non-Hermitian~\mycite{Luo2019PRL}, and superconducting~\mycite{LiuPRB2018, ZhuPRB2018, LaubscherPRR2019, YanPRL2018,WangPRL2018, KheirkhahPRB2020, YanPRL2019} systems.

In addition to these platforms, proposals to observe HOT physics in quantum spin systems~\mycite{dwivedi2018PRB,wang2021majorana, guo2020quantum} and solvable spin liquid models~\mycite{watanabe2021fractional,neupertPRB2018,dubinkin2019PRB} have also been put forward. While investigations of HOT phases in these spin systems have mainly focused on ground-state physics, their generalization to excited states is relatively less explored. A few recent studies have tried to bridge this gap by proposing spin models with magnetically ordered ground states. Although the ground states are topologically trivial, the HOT phases in these magnetic systems emerge as hinge~\mycite{mook2020chiral, park2021hinge} or corner~\mycite{li2019higher, silJOP2020, hirosawa2020magnonic} magnon excitations. In this paper, we explore the opposite regime; i.e., we study spin models without any magnetic order. We demonstrate that even spin-gapped quantum paramagnets, formed by coupled spin dimers, can host excitations that show HOT physics.

Spin dimers arise naturally in Mott insulators where certain adjacent pairs of sites are coupled by strong exchange interactions, with weaker inter-dimer exchange couplings~\mycite{shastry1981, Smith1989, sasago1997, tanaka2001, sakurai2002, ruegg2003, ruegg2005, giamarchi2008, sebastian2006, ronnow2017, rossPRL2019}. They can also arise in geometrically frustrated magnets which may exhibit spontaneous dimerization and lattice symmetry breaking via formation of valence bond crystals~\cite{majumdar1969,sandvik2007}. On a single dimer with strong antiferromagnetic intra-dimer Heisenberg exchange, the ground state is a singlet, and the gapped excited states form triplets with total spin $S=1$. Weaker inter-dimer couplings in the crystal induce hopping and pairing terms for the triplet excitations. Such singlet quantum paramagnets thus exhibit a spectrum of gapped and dispersive triplet excitations called \emph{triplons}.

With full spin-rotational symmetry, the triplons exhibit three decoupled and degenerate bands, corresponding to the three distinct spin states. However, in the presence of spin-orbit coupling (SOC) or an applied magnetic field, the different spin states get coupled and this band degeneracy is broken. It has been shown that the resulting triplon bands can exhibit non-trivial first order topology, with nonzero Chern numbers and a significant thermal Hall effect \cite{penc2015hall,coldea2017NPJ}. In our work, we show that such quantum paramagnets can also exhibit HOT phases and triplon corner-mode excitations that arise as near-mid-bandgap states in their triplet excitation spectrum.

This paper is organized as follows: In \Sec{sec:Models}, we formally define the two spin models which exhibit the HOT triplon phases. In \Sec{sec:triplons}, we develop the effective theory of triplon excitations for the spin models. The emergent triplon phases for the models, including the HOT phases, are discussed in \Sec{sec:phases}, and the symmetries that protect the HOT phases are discussed in \Sec{sec:symmetries}. We characterize the topological triplon phases using topological invariants such as the Wannier-sector polarizations in \Sec{sec:topo-inv}. Finally, we develop the effective edge theory for the models in \Sec{sec:edge-theo}, allowing us to show that the corner modes arise at the intersection between two edges. We conclude in \Sec{sec:conclusion} with a summary of the work and a discussion on possible experimental realization. Additional details about various results are provided in the appendices, and are referred to in the main text.
 
\section{Model Hamiltonians}\label{sec:Models}
\begin{figure*}
    \centering
\includegraphics[width=\textwidth]{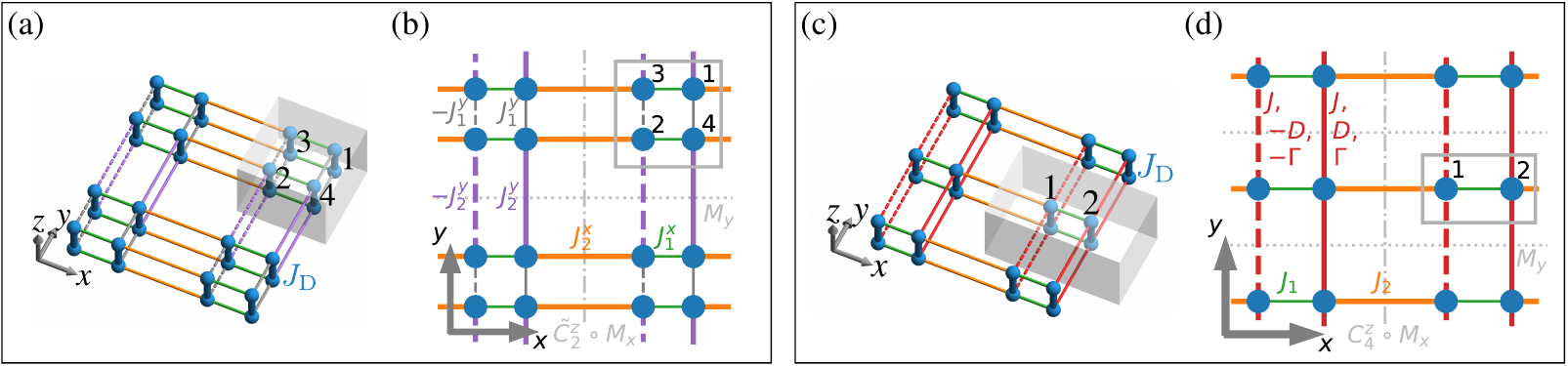}
\caption{Models: (a) The lattice of spin dimers hosting the triplon-BBH model (\modI), made up of a top (T) and a bottom (B) layer. The unit cell is composed of four dimers numbered 1 to 4. (b) A top-down view of the lattice in (a); thin (thick) horizontal bonds have $\Kh_1$ ($\Kh_2$) Heisenberg interactions, and thin (thick) vertical bonds have interactions proportional to $\Kv_1$ ($\Kv_2$), respectively. The sign of the interactions on the vertical bonds alternates as we move along the $x$-direction. The dashed vertical lines have $-\Kv_{1,2}$ and the solid vertical lines have $+\Kv_{1,2}$ interactions. (c) The bilayer spin dimer lattice used to realize a HOT phase in the \modII\ model for triplons with spin-orbit coupling (SOC). The model has a reduced unit cell (compared to (a)) consisting of two dimers numbered 1 and 2. (d) A top-down view of the lattice in (c); horizontal bonds have the same configuration as in (b) with thin (thick) bonds having $\K_1$ ($\K_2$) Heisenberg interactions. The vertical bonds have DM ($D$) and Gamma ($\Gamma$) interactions that introduce SOC into the model, in addition to a small Heisenberg interaction $\K$. The direction of DM and Gamma interactions alternates along the $x$ direction, with dashed lines having DM and Gamma interactions pointing along the $-\hat{z}$ direction, and solid lines in the $+\hat{z}$ direction. Mirror axes described in \Sec{sec:symmetries} are shown in the top-down views in dotted and dashed-dotted lines.
}\label{fig:models}
\end{figure*}

We consider an array of spin dimers described by the Hamiltonian
\begin{equation}\label{eqn:Hdim}
\Hdim= \J \sum_{\br} {\bf S}_{\br T} \cdot {\bf S}_{\br B},
\end{equation}
with $\J>0$, where spins live on a bilayer square lattice. 
The square-lattice sites are labeled by $\br$, and $T,B$ refer to `top' and `bottom' spins on the bilayer which form the dimer at $\br$. 
The unique ground state of the Hamiltonian $\Hdim$ is a direct product of interlayer dimer singlets at each site $\br$. 
The lowest energy excitations correspond to creating a triplet on any one of the dimers with the spin gap energy $\J$.
We next introduce two distinct models of inter-dimer exchange interactions and demonstrate that the resulting quantum paramagnets host HOT triplon excitations.

\subsection{Model I: The \modI~model}
Our first model, an XXZ model of inter-dimer couplings,
is directly inspired by the BBH model of spinless fermions \mycite{benalcazar2017quantized}. 
Hence, we name this the \modI~model, for ``triplon BBH model''.
To define the model, we group four dimers together into 
a unit cell as shown in \figs{fig:models}(a) and \ref{fig:models}(b).
The total spin Hamiltonian is given by 
\begin{subequations}\label{eqn:spin-ham-I}
\begin{gather}
    \HBBH = \Hdim + \HI 
    \\
    \HI = \!\!\! \sum_{\langle\br\br'\rangle,\l}\!\! \K^{\pdg}_{\br,\br'} \left[
        \zeta^{\pdg}_{\br,\br'} (\Sc{x}_{\br \l}\Sc{x}_{\br' \l} \!+\! \Sc{y}_{\br \l}\Sc{y}_{\br' \l}) + \Sc{z}_{\br \l}\Sc{z}_{\br'\l}
    \right]
\end{gather}
\end{subequations}
where $\l=T,B$ labels the layers. The inter-dimer exchange
$\K_{\br,\br'}$ takes on values $\K_1$ and $\K_2$
for neighboring sites which are in the same unit cell or
in adjacent unit cells, respectively. The coefficient $\zeta_{\br,\br'}$ defines the sign of the XY exchange, taking
on values $+1$ $(-1)$ on the solid (dashed) bonds as shown in
\fig{fig:models}(b); as we show below, this sign choice
leads to the triplons sensing a $\pi$-flux~\footnote{This is not a physical electromagnetic flux, but rather the phase accumulated by the triplon wavefunction as the triplon hops around a single plaquette.} in each 
elementary square plaquette. For this Hamiltonian, only
two of the three triplons ($x,y$) will form HOT bands, as we discuss later,
while the $z$-triplon remains in a topologically trivial phase.
The model may be considered an example of Su-Schrieffer-Heeger (SSH)~\mycite{SSH1979} spin ladders running along $\hat{x}$ which are coupled by frustrated inter-ladder interactions; a recent experimental example of such an SSH spin ladder is discussed in Ref.~\onlinecite{nawa2019triplon}.

\subsection{Model II: The \modII~model}
To get higher-order topology to emerge in 2D fermionic systems, it has been shown that we need at least four degrees of freedom per unit cell \mycite{benalcazar2017quantized}.
Heuristically, the polarizations defining the first-order topologies of two bands can then potentially cancel with each other, allowing for the higher order quadrupole moment to emerge as the proper characterization of the nontrivial band topology of this band pair. 
In our first model, the different triplon flavors are decoupled, and the relevant degrees of freedom arise from having four dimers in each unit cell. 
In our second model below, we instead include SOC that mixes two of the three triplon flavors, which permits a smaller unit cell 
consisting of two dimers. Therefore, we refer to this model as the \modII\ model, short for ``triplons with spin-orbit coupling''.

The model is constructed (see \figs{fig:models}(c) and (d)) as an array of 1D ladders along $y$, each ladder being described by a model which hosts triplon bands with nontrivial \emph{first}-order topology as shown by Joshi and Schnyder~\mycite{joshi2017topological}.
In the transverse ($x$) direction, the chains are coupled by alternating Heisenberg couplings
which realize an SSH spin model. The Hamiltonian for the \modII\ model is given by
\begin{subequations}\label{eqn:spin-ham-II}
\begin{align}
\HJS = \Hdim + \HII
\end{align}
\begin{multline}
    \HII \! = \!\!\!\! \sum_{\langle\br\br'\rangle,\l} 
        \!\!\! \K^{\pdg}_{\br,\br'} {\bf S}^\pdg_{\br \l} \cdot {\bf S}^\pdg_{\br'\l}
    + D \!\!\!\!\! \sum_{\langle\br\br'\rangle_y,\l} 
        \!\!\!\! \zeta_{\br,\br'} (S^x_{\br\l} S^y_{\br+y \l} - S^y_{\br\l} S^x_{\br+y \l})
    \\
    + \Gamma \sum_{\langle\br\br'\rangle_y,\l} 
        \zeta_{\br,\br'} (S^x_{\br\l} S^y_{\br+y \l} + S^y_{\br\l} S^x_{\br+y \l})
\end{multline}
\end{subequations}
Along the $x$ direction, $\K_{\br,\br'}$ takes on two distinct values $\K_1$ and $\K_2$ for intra-unit-cell nearest-neighbor spins and inter-unit-cell nearest neighbor spins, respectively (see \fig{fig:models}(d)). 
Along the $y$ direction, $\K_{\br,\br'}=\K$ for nearest-neighbor spins.
The $D$ and $\Gamma$ terms represent additional off-diagonal exchange interactions---the antisymmetric Dzyaloshinskii-Moriya (DM) coupling $D$ and its symmetric counterpart $\Gamma$. 
As in the previous model, the coefficient $\zeta_{\br,\br'}$ takes on values $+1$ $(-1)$ on the solid (dashed) bonds as shown in \fig{fig:models}(d), so that the DM and $\Gamma$ interactions alternate in sign from one $y$-direction ladder to the next. 
\section{Triplon Excitations}\label{sec:triplons}

The models defined in \Sec{sec:Models} possess a trivial quantum paramagnetic ground state when the inter-dimer exchanges are much weaker than $\J$ (\eqn{eqn:Hdim}). 
In order to describe the excitations above the singlet ground state, we use the bond operator formalism~\mycite{sachdev1990bondoperator, collins2006modified}, which maps the triplets and singlet states of each dimer to bosons as follows
\begin{equation}
\begin{split}
    \td_x\vac &= \hphantom{-} \ci (\ket{\!\ua\ua}-\ket{\!\da\da})/\sqrt{2}
    \\
    \td_y\vac &= \hphantom{-\ci}(\ket{\!\ua\ua}+\ket{\!\da\da})/\sqrt{2}
    \\
    \td_z\vac &= -\ci (\ket{\!\ua\da}+\ket{\!\da\ua})/\sqrt{2}
    \\
    \sd\vac &= \hphantom{-\ci} \singlet\ .
\end{split}
\end{equation}
Here we have labeled the three triplet flavors~\mycite{penc2011PRB} as $x,y,z$. The bosonic operators $\td_\alpha$ (with $\alpha \in \{x,y,z\}$) create triplons from the vacuum state $\vac$, and $\sd$ does the same for a singlet. Since each dimer has exactly one singlet or triplet, we must impose the constraint $\sd s + \sum_\alpha \td_\alpha t^\pdg_\alpha=1$, where $\sd s$ and $\td_\alpha t^\pdg_\alpha$ gives the number densities for singlets and $\alpha$ triplets respectively. With this constraint, the 
spin operators on the two sites at a dimer can be represented 
in terms of the bosons as follows:
\begin{subequations}\label{eq:spin_triplon_decomposition}
\begin{align}
    S^{\alpha}_{\br T} &= \frac{1}{2} \Bigl( +\ci \td_{\br\alpha}s^\pdg_{\br} - \ci \sd_{\br} t^\pdg_{\br\alpha} -\ci \varepsilon^\pdg_{\alpha\beta\gamma}\td_{\br\beta} t^\pdg_{\br\gamma} \Bigr),
    \\
    S^{\alpha}_{\br B} &= \frac{1}{2} \Bigl( -\ci \td_{\br\alpha}s^\pdg_{\br} + \ci \sd_{\br} t^\pdg_{\br\alpha} -\ci \varepsilon^\pdg_{\alpha\beta\gamma}\td_{\br\beta} 
    t^\pdg_{\br\gamma} \Bigr).
\end{align}
\end{subequations}
Deep in the paramagnetic phase, the singlet density is approximately $1$.
Throughout this paper we will work in this limit, 
and therefore set $s$ and $\sd$ to the c-number 1 \cite{sachdev1990bondoperator}. Furthermore, we 
shall only keep terms that are bilinear in the triplon operators $\td_\alpha$ and $t^\pdg_\alpha$~\mycite{sachdev1990bondoperator, joshi2017topological, penc2015hall}.

In deriving the effective triplon Hamiltonians below to describe the spin-gapped excitations, we relabel the sites $\br := (\site{\R}{a})$ where $\bR$ denotes unit cell and $a$ stands for the sublattice index. 
For the \modI\ model (model I), the sublattice index $a \in \{1,2,3,4\}$, while for the \modII\ model (model II), $a \in \{1,2\}$. 
For compactness of notation, we define two-component Nambu spinors 
$\phi^\dg_{\site{\R}{a}\alpha} = [t^\dg_{\site{\R}{a}\alpha} ~ t^\pdg_{\site{\R}{a}\alpha}]$.
For both models, the $z$ triplon remains decoupled with topologically trivial bands. We will 
henceforth focus  {\it only} on those sectors which exhibit nontrivial topology; with this understanding, 
the triplon label $\alpha$ will be used to denote {\it only} $\alpha \in \{x,y\}$. 

\paragraph{Model I (\modI\ model)}
In the first model, the different triplet flavor states $\alpha$ are decoupled at quadratic order due to the absence of SOC and Zeeman fields. We construct the eight-component bosonic Nambu \spinor\ $\Phi^\dg_{\R\alpha} = [\phi^\dg_{\site{\R}{1}\alpha}~\phi^\dg_{\site{\R}{2}\alpha}~\phi^\dg_{\site{\R}{3}\alpha}~\phi^\dg_{\site{\R}{4}\alpha}]$.
The real-space triplon Hamiltonian, obtained by expressing the spin operators in \eqn{eqn:spin-ham-I} using the relations in \eqn{eq:spin_triplon_decomposition}, is then given by
\begin{subequations}\label{eqn:trip-ham-real-I}
\begin{equation}
    \opHtripI = \frac{1}{2} \sum_{\R, \boldsymbol{\delta}, \alpha} \Fid_{\R+\boldsymbol{\delta}\alpha} \, \M^\pdg_{\boldsymbol{\delta}} \, \Fi_{\R\alpha},
\end{equation}
with the coupling matrices
\begin{equation}
\begin{split}
    \M_0 &= \J\sigma_0\eta_0\tau_0 \!+\! \left( \! \frac{\K_1}{2} \sigma_1\eta_0 \!-\! \frac{\K_1}{2} \sigma_2\eta_2 \!\right) \! \left(\tau_0 \!-\! \tau_1\right)
    \\
    \M_{+\hat{x}} &= \M^\dagger_{-\hat{x}} = \frac{\K_2}{4} \left(\sigma_1\eta_0 - \ci\sigma_2\eta_3 \right) \left(\tau_0 \!-\! \tau_1\right)
    \\
    \M_{+\hat{y}} &= \M^\dagger_{-\hat{y}} = - \frac{\K_2}{4} \ci\sigma_2 \left(\eta_1-\ci\eta_2\right) \left(\tau_0 \!-\! \tau_1\right).
\end{split}
\end{equation}
\end{subequations}
Here, $\sigma_i$, $\eta_i$, and $\tau_i$ correspond to sets of Pauli matrices with Kronecker product implied between them. Specifically, $\tau_i$ acts in Nambu 
(particle-hole) space, while $\sigma_i$ and $\eta_i$ act in the space of sublattices $a\in\{1,2,3,4\}$. In terms of the $(\sigma_3,\eta_3)$ eigenvalues, the sublattices $a=(1,2,3,4)$ correspond respectively to $(++,+-,-+,--)$.
Fourier-transforming this yields the momentum-space Hamiltonian 
$\opHtripI \!=\! \frac{1}{2} \sum_{\k\alpha} \Fid_{\k\alpha} H_{\k}^\text{tBBH} \Fi_{\k\alpha}$
~\footnote{The redundancy intrinsic to the Nambu formalism is eliminated by choosing coefficient matrices to be particle-hole symmetric. In momentum space, this constraint takes the form $H_{\k} = \tau_1 H_{-\k}^* \tau_1$}, where $H_{\k}^\text{tBBH}$ is an $8\times 8$ matrix independent of $\alpha$, given by
\begin{multline}\label{eqn:trip-bloch-ham-model-I}
    H_{\k}^\text{tBBH} = \J\sigma_0\eta_0\tau_0 \!+\! \left(\! \frac{\K_1}{2} \sigma_1\eta_0 \!-\! \frac{\K_1}{2} \sigma_2\eta_2 \!\right) \! \left(\! \tau_0 \!-\! \tau_1 \!\right)
    \\
    + \frac{\K_2}{2} \left( \cos k_x \sigma_1\eta_0 \!-\! \sin k_x \sigma_2\eta_3 \right) \! \left(\! \tau_0 \!-\! \tau_1 \!\right)
    \\
    - \frac{\K_2}{2} \sigma_2 \left( \cos k_y \eta_2 + \sin k_y \eta_1 \right) \left(\! \tau_0 \!-\! \tau_1 \!\right).
\end{multline}

\paragraph{Model II (\modII\ model)}
In the second model, SOC couples the $x$ and $y$ triplon sectors.
We define the eight-component bosonic Nambu \spinor\ $\Psi^\dg_{\R} = [\phi^\dg_{\site{\R}{1}x}~\phi^\dg_{\site{\R}{1}y}~\phi^\dg_{\site{\R}{2}x}~\phi^\dg_{\site{\R}{2}y}]$, in terms of which the triplon Hamiltonian is given by
\begin{subequations}\label{eqn:trip-ham-real-II}
\begin{equation}
    \opHtripII = \frac{1}{2} \sum_{\R, \boldsymbol{\delta}} \Sid_{\R+\boldsymbol{\delta}} \M^\pdg_{\boldsymbol{\delta}} \Si^\pdg_\R,
\end{equation}
with the coupling matrices given by
\begin{equation}
\begin{split}
    \M_0 &= \J \sigma_0 \eta_0 \tau_0 
+ \frac{\K_1}{2} \sigma_1 \eta_0 (\tau_0 \!-\! \tau_1)
    \\
    \M_{+\hat{x}} &\!=\! \M_{-\hat{x}}^\dagger \!=\! \frac{\K_2}{4} \left( \sigma_1 + \ci \sigma_2 \right) \eta_0 \left( \tau_0 \!-\! \tau_1 \right)
    \\
    \M_{+\hat{y}} &\!=\! \M_{-\hat{y}}^\dagger \!=\! \frac{\K \sigma_0\eta_0 + \sigma_3 (\Gamma \eta_1 \!+\! \ci D\eta_2)}{2} \! \left(\tau_0 \!-\! \tau_1\right).
\end{split}
\end{equation}
\end{subequations}
While $\tau_i$ Pauli matrices again act in Nambu (particle-hole) space, here the $\sigma_i$ and $\eta_i$ Pauli matrices respectively act on the sublattice index $a \in (1,2)$ and triplon flavor index $\alpha \in (x,y)$. 
Thus the two triplon flavors here take on the role of the two additional sublattices in model I.
Fourier-transforming leads to the $8\times 8$ Bloch Hamiltonian:
\begin{multline}\label{eqn:trip-bloch-ham-model-II}
H_{\k}^\text{tSOC} \!\!=\! \J \sigma_0 \eta_0 \tau_0 
+ \frac{\K_1}{2} \sigma_1 \eta_0 (\! \tau_0 \!-\! \tau_1 \!)
    \\
    \! \!+\! \frac{\K_2}{2} \left( \cos\! k_x \sigma_1 \!+\! \sin\! k_x \sigma_2 \right) \eta_0 (\! \tau_0 \!-\! \tau_1 \!)
    \\
    \!\!+\! \bigl( 
      (\! \K\sigma_0\eta_0 \!+\! \Gamma\sigma_3\eta_1 \!)\! \cos\! k_y \!+\!\! D \sin\! k_y \sigma_3 \eta_2
    \bigr) (\! \tau_0 \!-\! \tau_1 \!).
\end{multline}

\paragraph{Diagonalization} Triplon Hamiltonians like $\opHtripI$ and $\opHtripII$ are in general bosonic Bogoliubov--de Gennes (BdG) Hamiltonians~\mycite{sachdev1990bondoperator}. We briefly review the process for solving such Hamiltonians: writing $\hat{H} = \frac{1}{2} \sum_{\k} \Phi_\k^\dagger H_\k \Phi_\k$, $\hat{H}$ is diagonalized via a paraunitary Bogoliubov transformation $\Phi_\k = T_\k \Gamma_\k$, where the matrices $T_\k$ satisfy $T_\k^\dagger \tau_3 T_\k = \tau_3$, where $\tau_3$, short for $1_n \otimes \tau_3$, is the third Pauli matrix in Nambu space~\mycite{COLPA1978327, blaizot1986quantum, KAWAGUCHI2012253, PhysRevB.100.075414}. Provided $H_\k$ is positive definite, such a transformation $T_\k$ exists, and the eigenvectors of $\tau_3H_\k$ can be chosen to construct the columns of $T_\k$. The BdG structure of the Hamiltonian guarantees that half the eigenvalues of $\tau_3H_\k$ are positive (particle like) and the other half is negative (hole like). The positive eigenvalues give the energies for the bosonic modes at a given momentum $\k$. The diagonalization procedure for a position-space BdG Hamiltonian can also be performed in a similar fashion.

\section{HOT phase and Topological transition}\label{sec:phases}
\begin{figure*}
    \centering
\includegraphics[scale=1.]{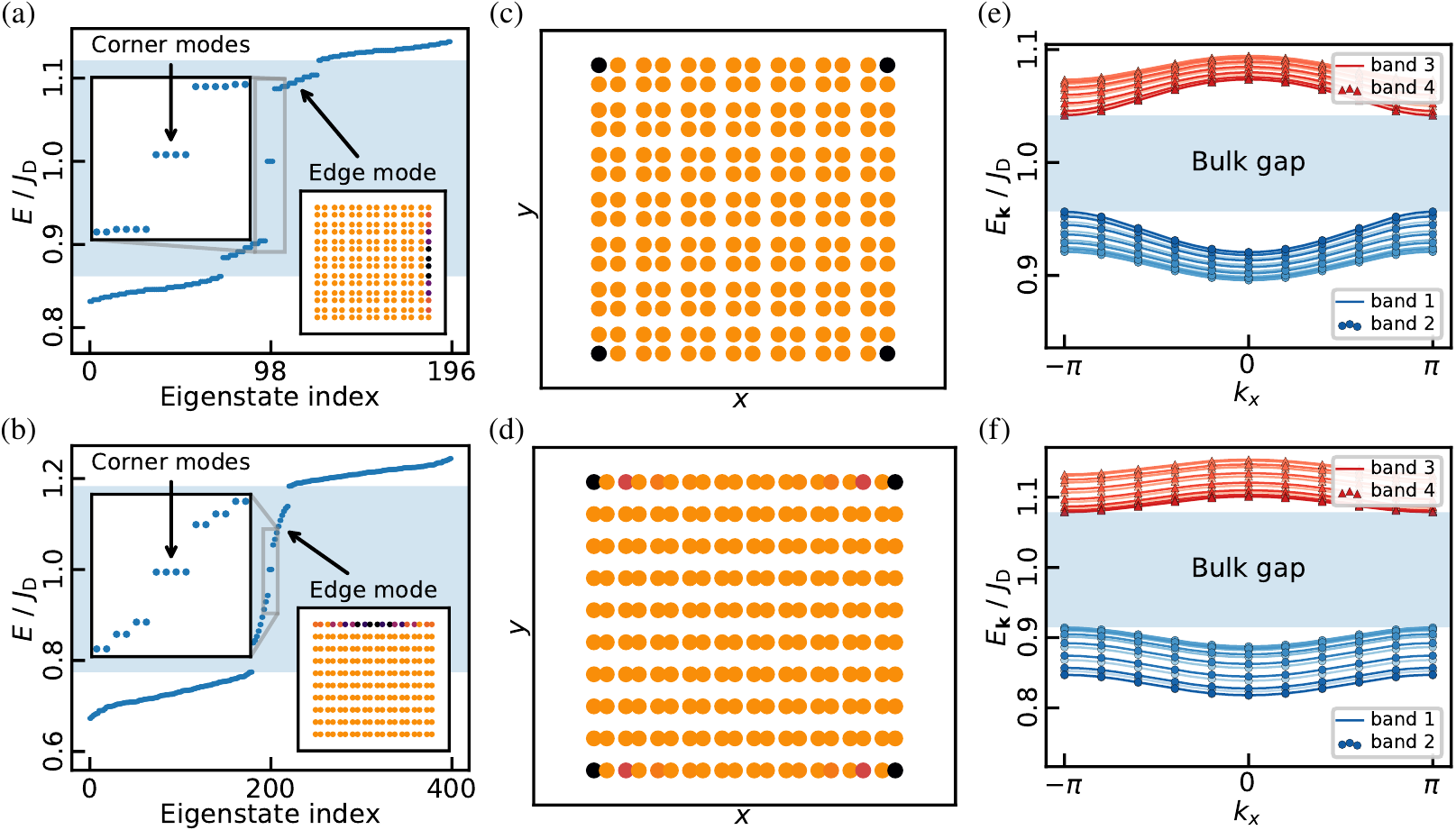}
    \caption{Corner modes and Band dispersions: (a) Energy spectrum for the \modI\ model with open boundary conditions (OBC), showing four corner modes (top inset) and edge modes inside the bulk bandgap (shaded region). The corner modes are pinned to the near-midgap energy $E/\J\approx 1$. The triplon density for a typical edge mode is given in the bottom-right inset. (b) Energy spectrum for the \modII~model in the HOT phase, with OBC, showing corner modes and edge modes like in (a). Figures (c) and (d) give the average triplon density profiles over all four corner modes of the \modI~and \modII~models, respectively. The triplon excitations are localized at the corners of the lattice (see \fig{fig:models}(b), (d)). (e)  and (f) show the band dispersions (orthographically projected onto the $k_y = 0$ plane), and the bulk bandgap, for the triplon modes in the \modI\ and \modII\ models, respectively. The parameter values used for plotting the dispersions in (e) and (f) are $\Kh_1/\J=\Kv_1/\J=0.04$, $\Kh_2/\J=\Kv_2/\J=0.1$ and  $\K_1/\J=0.06$, $\K_2/\J=0.1$, $\K/\J=0.03$, $\Gamma/\J=0.11$, $D/\J=0.1$, respectively. The triplon dispersion for both the models comprises four Bloch bands, out of which bands 1 (lines) and 2 ($\circ$) form a degenerate pair and bands 3 (lines) and 4 ({\tiny$\triangle$}) form another.
}
    \label{fig:corner-modes}\label{fig:band-structures}
\end{figure*}

A characteristic feature of a 2D HOT phase is the existence of topological corner modes. 
To check for these corner modes, we solve the triplon real space Hamiltonians in \eqn{eqn:trip-ham-real-I} and \eqn{eqn:trip-ham-real-II} on a finite geometry with open boundary conditions (OBC) in the $x$ and $y$ directions.

For the first model, given by $\opHtripI$ (\eqn{eqn:trip-ham-real-I}), we fix parameters $\K_1/\J = 0.02$ and $\K_2/\J = 0.2$ and plot the energy spectrum for the $t_x$ triplons in \fig{fig:corner-modes}(a). It shows an energy bandgap with four modes near the middle of the bandgap ($E/\J\approx1.0$), right about where we would expect to find the corner-mode energies. We use the term ``near" to specify the location of these energies, as we find the spectrum is \emph{not} symmetric about $E/\J=1.0$ for a finite value of $\J$. However, the spectrum becomes symmetric asymptotically as $\J\to\infty$, and the energies corresponding to these four modes shift to exactly the middle of the bandgap. To verify that these are corner modes, we plot the real-space triplon density averaged over all four of these modes and overlay it over our finite lattice geometry; see \fig{fig:corner-modes}(c). Indeed, we see that the density peaks at the corners, confirming the existence of corner modes. In addition to the corner modes, we find \emph{edge} modes whose energies lie inside the bandgap but are not pinned to the energy $E\approx\J$ inside the bandgap (see \fig{fig:corner-modes}(a))~\footnote{In fact, the edge mode energies can be pushed into the bulk spectrum by tuning the parameters appropriately without destroying the corner modes.}; such states also exist in the fermionic BBH model~\mycite{benalcazar2017quantized}. We show the density profile of one of these edge modes in the bottom inset of \fig{fig:corner-modes}(a).

To check for corner modes in the second model, we use parameter values $\K_1/\J = 0.1$, $\K_2/\J = 0.2$, $D/\J = 0.2$, $\Gamma/\J = 0.22$, and $\K/\J = 0.01$ in $\HtripII$ (\eqn{eqn:trip-ham-real-II}). We plot the energy spectrum in \fig{fig:corner-modes}(b) and once again find four modes pinned near the center of the bandgap at $E/\J\approx1.0$. Like before, a plot of the real-space triplon density profile averaged over the four near-midgap states (\fig{fig:corner-modes}(d)) shows that they reside at the corners and decay rapidly into the bulk, verifying that they are indeed corner modes.
Also similar to the first model, we find edge modes with energies considerably away from the middle of (but inside) the bandgap for the second model. The density profile for a typical edge mode is given in \fig{fig:corner-modes}(b) bottom inset.

\paragraph*{Band structures:}
The band structures for the two models (\eqns{eqn:trip-bloch-ham-model-I} and \ref{eqn:trip-bloch-ham-model-II}) are shown in \fig{fig:band-structures}(e) and (f), with respective parameter values discussed in the caption. 
In each model, the particle-like bosonic modes are made up of four bands (see the last paragraph of \Sec{sec:triplons}). 
In the first model, the lowest two energy bands, i.e., bands $(1,2)$, are degenerate, and so are bands $(3,4)$; see \fig{fig:band-structures}(e). 
The bulk bandgap occurs between band pair $(1,2)$ and band pair $(3,4)$. 
The corner modes (see \fig{fig:corner-modes}(a)) appear at an energy $E$ near the middle of this bandgap with $E\approx \J$, where the intra-dimer Heisenberg exchange $\J$ is the dominant energy scale. 
The bands for the second model are plotted in \fig{fig:band-structures}(f), and have the same type of degeneracies as the first model. 
Other features are also qualitatively similar to the first model, e.g., the bulk bandgap appearing between bands 2 and 3. 
Further details and explicit expressions for the band dispersions are discussed in \app{app:bandstructures}.

\subsection{Phase diagrams}

Having established the presence of corner modes for a set of parameter values, we now explore the full phase diagram for the two models. In particular, we look for ways to drive transitions from the HOT triplon phase to other phases with trivial (or possibly first-order) topology.

\subsubsection{Model I: \modI~model}
\begin{figure*}
    \centering
\includegraphics[width=\textwidth]{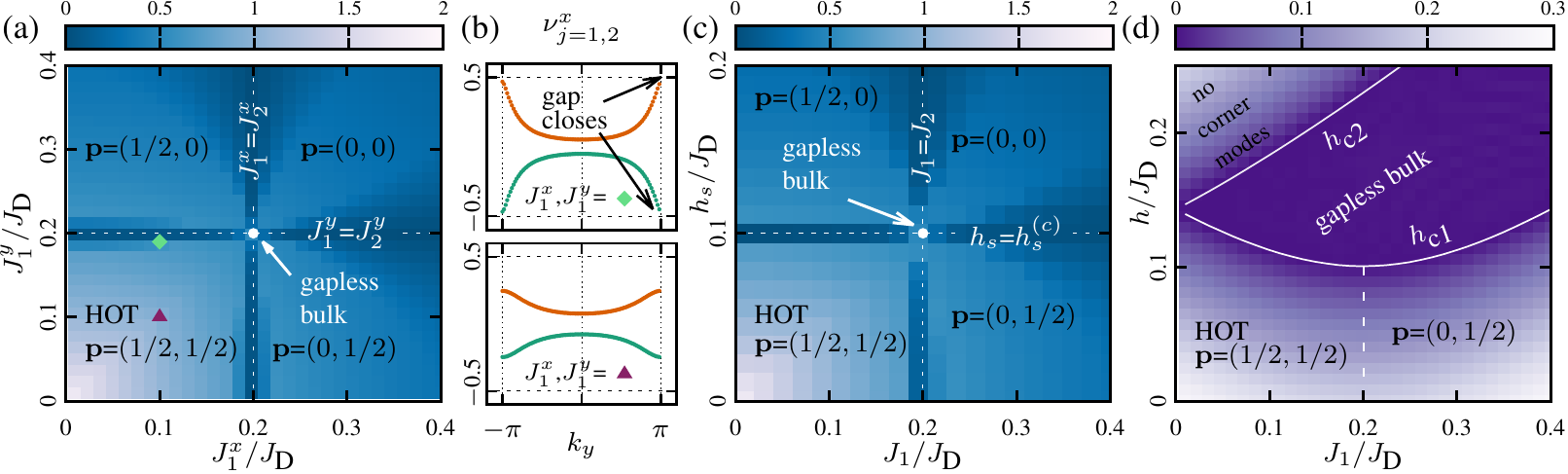}
    \caption{Phase diagrams: 
    (a) The phases of the BBH model for triplons (\modI); the HOT-triplon phase appears when $\Khv_{1}<\Khv_{2}$. 
    The Wannier bandgap (see \Sec{sec:topo-inv}), shown with a color-density plot, closes along the dashed lines $\Kh_1=\Kh_2$, $\Kv_1=\Kv_2$. 
    The bulk bandgap closes only at the intersection point of the lines. 
    The Wannier-sector polarizations $\p=(p_x,p_y)$ are quantized to either $0$ or $1/2$ (see \Sec{sec:topo-inv}).
    (b) Gapless (top inset) and gapped (bottom inset) Wannier bands $\WBc{x}(\k)$ appearing in phases close to the horizontal-transition line ($\diamond$) and inside the HOT region ({\small$\triangle$}) in panel (a), respectively. 
    (c) Phases of the second model with SOC triplons (\modII) in a staggered field $\hs$. 
    The phases are in one-to-one correspondence with the \modI~model in panel (a); the staggered-field strength $\hs$ acts like $\Kv_1$ from panel (a). 
    (d) Phase diagram for the \modII~model in a homogeneous field $h$; a HOT phase is found when $h<\hchomI$ and $\K_1<\K_2$. 
    The Wannier-sector polarizations $\p$ are quantized for $h<\hchomI$. 
    The bulk bandgap (shown with a color-density plot) closes when $\hchomI\leq h \leq \hchomII$, and the system cannot harbor protected near-midgap modes. 
    Beyond $h>\hchomII$, the system is gapped but does not have any corner modes.
    In panels (c) and (d), the parameter values $D/\J = \Gamma/\J = 0.1$ are used.
    }
    \label{fig:phase-diagrams}
\end{figure*}

In the fermionic BBH model, the complete phase diagram for the topological phases is accessed by allowing the hopping amplitudes on the 2D square lattice to be anisotropic; in other words, the amplitudes along the $x$ direction are tuned independently from those along the $y$ direction.
Since our first model closely resembles the fermionic BBH model, we introduce a similar anisotropy into the tBBH model by making the Heisenberg exchange interactions along the horizontal ($x$-direction) bonds independent from those along vertical ($y$-direction) ones; see \fig{fig:models}(b). 
This entails that the triplon Hamiltonian in \eqn{eqn:trip-bloch-ham-model-I} is redefined as
\begin{multline}\label{eqn:trip-bloch-ham-model-I-aniso}
    H_{\k}^\text{tBBH} \overset{\text{redef.}}{=} \J\sigma_0\eta_0\tau_0 \!+\! \left(\! \frac{\Kh_1}{2} \sigma_1\eta_0 \!-\! \frac{\Kv_1}{2} \sigma_2\eta_2 \!\right) \! \left(\! \tau_0 \!-\! \tau_1 \!\right)
    \\
    + \frac{\Kh_2}{2} \left( \cos k_x \sigma_1\eta_0 \!-\! \sin k_x \sigma_2\eta_3 \right) \! \left(\! \tau_0 \!-\! \tau_1 \!\right)
    \\
    - \frac{\Kv_2}{2} \sigma_2 \left( \cos k_y \eta_2 + \sin k_y \eta_1 \right) \left(\! \tau_0 \!-\! \tau_1 \!\right),
\end{multline}
where $\Kh_{1,2}$ and $\Kv_{1,2}$ represent the strengths of interactions on the $x$ and $y$ bonds, respectively. 
Continuing with the analogy between the tBBH and fermionic BBH models, we expect to drive a transition from the HOT phase to other phases by tuning the ratios $\Kh_1/\Kh_2$ and $\Kv_1/\Kv_2$.
Setting $\Kh_2/\J=\Kv_2/\J=0.2$ in $H_{\k}^\text{tBBH}$ (\eqn{eqn:trip-bloch-ham-model-I-aniso}), we construct the phase diagram in \fig{fig:phase-diagrams}(a) by separately tuning $\Kh_1/\J$ and $\Kv_1/\J$ from $0.0$ to $0.4$. 
The HOT triplon phase occurs in the region with both $\Kh_1<\Kh_2$ \emph{and} $\Kv_1<\Kv_2$. 
At any point in this region of parameter space, the system hosts stable corner modes. 
If either $\Kh_1/\Kh_2$ or $\Kv_1/\Kv_2$ are made greater than $1.0$, the HOT phase is destroyed and the corner modes disappear. 
Notably, the phase boundary of the transition from a HOT to a non-HOT phase, indicated by the dashed lines $\Khv_1=\Khv_2$ in \fig{fig:phase-diagrams}(a), does not coincide with a closing of the bulk bandgap! 
This unique feature of HOT phases underlines that they are truly distinct from conventional first-order topological phases. 
In the phase diagram of \fig{fig:phase-diagrams}(a), the bulk bandgap (\fig{fig:band-structures}(e)), given by the formula
\begin{equation}\label{eq:modIbulkgap}
\begin{split}
       \Blkgap &= E_+^\text{(min)} - E_-^\text{(max)}
       \\
       &= \frac{ 2\J \sqrt{ \bigl(|\Kh_1|-|\Kh_2|\bigr)^2+ \bigl(|\Kv_1|-|\Kv_2|\bigr)^2 } }{ E_+^\text{(min)} + E_-^\text{(max)} }\ ,
\end{split}
\end{equation}
closes only at a single point when $\Kh_1/\Kh_2=\Kv_1/\Kv_2=1$ (see \fig{fig:phase-diagrams}(a)). In \eqn{eq:modIbulkgap}, $E_+^\text{(min)}$ and $E_-^\text{(max)}$ (where $E_+^\text{(min)} \geq E_-^\text{(max)} > 0$) are the maximum and minimum values for the bands 1,2 and 3,4, respectively\ifdefined\useapp
---explicit expressions are provided in \app{app:bandstructures}\fi. Though the HOT phase boundaries are not marked by a closing of the bulk bandgap, they do coincide with bandgap closings in the \emph{edge theories} of the model---this is further discussed in \Sec{sec:topo-inv} together with the topological invariants for our models.

\subsubsection{Model II: \modII~model}
Moving on to constructing a phase diagram for our second model, the tuning parameters at our disposal to drive a transition are $\K_1$, $\K_2$, $D$ and $\Gamma$. We set $\K$, the Heisenberg exchange on vertical bonds, to zero for the present discussion. Since the interactions along the $x$ direction are the same as in the first model, we once again choose the ratio $\K_1/\K_2$ as one of the parameters. Interestingly, tuning the strengths of $D$ or $\Gamma$ (as long as $D,\ \Gamma\ll \J$) does not destroy the HOT-triplon phase! We have checked that the corner modes persist as long as both $D$, $\Gamma$ are not equal to \emph{zero}.
Therefore, we must introduce additional terms in the Hamiltonian (\eqn{eqn:spin-ham-II}) to drive a transition. We explore two such avenues: (a) by applying a \emph{staggered} Zeeman field
\begin{align}\label{eqn:hs-def}
    \hat{H}_{\textup{stag-}B}^\text{spin} = - \hs\sum_{\R,a=1,2,l}(-1)^a\Sc{z}_{\site{\R}{a},l}
\end{align}
of strength $\hs$, or (b) by applying a homogeneous Zeeman field
\begin{align}\label{eqn:h-def}
    \hat{H}_{\textup{hom-}B}^\text{spin} = - h\sum_{\R,a=1,2,l}\Sc{z}_{\site{\R}{a},l},
\end{align}
controlled by the strength $h$. In the staggered field configuration, the Zeeman field at dimers 1 and 2 of the unit cell respectively points in the $-\hat{z}$ and $+\hat{z}$ directions, while for the homogeneous configuration it points everywhere in the $+\hat{z}$ direction (see \fig{fig:models}(d)). The corresponding spin Hamiltonian in each case takes the form $\HBBHII + \hat{H}_{\textup{stag-}B}$ and $\HBBHII + \hat{H}_{\textup{hom-}B}$, respectively; the triplon Hamiltonians are discussed in \app{app:bandstructures}. We discuss the phase diagrams for the staggered and homogeneous field configurations separately.

\paragraph{Staggered field}
We begin with the staggered field, since the associated phase diagram will turn out to be analogous to that of the \modI\ model. To investigate the phase diagram, we set $D/\J=\Gamma/\J=0.1$ and tune $\hs/\J$ (the staggered-field strength) from $0.0$ to $0.2$. Like for the first model, we tune $\K_1/\J$ from $0.0$ to $0.4$ with $\K_2/\J=0.2$. We show the resulting phases in \fig{fig:phase-diagrams}(b). Indeed, we find that the HOT phase transitions to a different topological state when $\hs$ becomes greater than a critical value $\hcstag\approx D$. Also, like in the first model, the HOT-phase disappears when $\K_1>\K_2$. Both these transitions are in general \emph{not} marked by a bulk bandgap (\fig{fig:band-structures}(f)) closing except at the point $\K_1=\K_2$ and 
\begin{align}\label{eqn:hcstag}
\hs = \hcstag := \frac{\sqrt{\J^2+2 D \J}-\sqrt{\J^2-2 D \J}}{2};
\end{align}
see \fig{fig:phase-diagrams}(b); see also \app{app:bandstructures} for the derivation of \eqn{eqn:hcstag}.

\paragraph{Homogeneous field}
Keeping all other parameters unchanged, we set $\hs = 0$ and tune $h/\J$ from $0.0$ to $2.0$. The resulting phase diagram, shown in \fig{fig:phase-diagrams}(c), is more complex than those encountered so far. There are now two critical field strengths $\hchomI$ and $\hchomII$, given by
\begin{equation}
\begin{split}
    \hchomI(\K_1) &= \left(\sqrt{\J^2+\J f_-(\K_1)}-\sqrt{\J^2-\J f_-(\K_1)}\right) \Big/ 2
    \\
    \hchomII(\K_1) &= \left(\sqrt{\J^2+\J f_+(\K_1)}-\sqrt{\J^2-\J f_+(\K_1)}\right) \Big/ 2
    \\
    f_-(\K_1) &= \sqrt{4 D^2+(|\K_1|-|\K_2|)^2}
    \\
    f_+(\K_1) &= \sqrt{4 D^2+(|\K_1|+|\K_2|)^2},
\end{split}
\end{equation}
that separate the various phases; see \app{app:bandstructures} for derivations. The HOT triplon phase persists as long as $h<\hchomI$ and $\K_1<\K_2$. Similar to the previous cases, the system transitions from a HOT phase to one with \emph{no} corner modes when $\K_1$ is tuned past $\K_2$, with the bulk bandgap remaining open in the process. However, for $\hchomI<h<\hchomII$, the bulk bandgap is absent, thereby precluding the existence of protected near-midgap states of any order (edge states or corner states). Beyond $h>\hchomII$, the bulk bandgap reopens but the HOT phase is not recovered for any values of $\K_1$ and $\K_2$.

The phases encountered in the above phase diagrams, especially those without robust corner modes, can be further characterized using topological invariants. We develop this characterization in more detail in \Sec{sec:topo-inv} on topological invariants.
 
\section{Symmetries}\label{sec:symmetries}
Usually, a set of non-commuting symmetries protect the HOT phases hosting corner modes~\mycite{benalcazar2017quantized}. For example, in the two-dimensional fermionic BBH model, there exists a pair of mirror-like symmetries $\Mx$ (reflection about the $y$ axis) and $\My$ (reflection about the $x$ axis) that \emph{anti-commute}~\mycite{benalcazar2017quantized}. 
As we shall demonstrate below, like in the BBH model, usual (commuting) mirror symmetries preclude the existence of an HOT phase in the spin-dimer models we consider, whereas non-commuting mirror symmetries allow for HOT.
We now discuss analogues of these non-commuting mirror-like symmetries for our two models below.

\subsection{Model I: \modI~model} 
For the first model, we find that the canonical mirror operation $M_x$, i.e., reflection about a line parallel to the $y$ axis and passing through the center of a unit cell (\fig{fig:models}(b)), is \emph{not} a symmetry by itself. Therefore, following the mirror reflection, we perform the operation $\tilde{C}^z_2$, a $\pi$ spin rotation about the $z$ axis acting \emph{only} on the spins of dimers 1 and 3 (\fig{fig:models}(b)), making them transform as
\begin{subequations}
\begin{equation}
    \begin{bmatrix}
        \Sc{x}_{\site{\R}{(1,3)},\l} \\ \Sc{y}_{\site{\R}{(1,3)},\l} \\ \Sc{z}_{\site{\R}{(1,3)},\l}
    \end{bmatrix}
    \!\!\overset{M_x}{\rightarrow}\!\!
    \begin{bmatrix}
        \hphantom{-} \Sc{x}_{\site{\R'}{(3,1)},\l} \\ -\Sc{y}_{\site{\R'}{(3,1)},\l} \\ -\Sc{z}_{\site{\R'}{(3,1)},\l}
    \end{bmatrix}
    \!\!\overset{\tilde{C}^z_2}{\rightarrow}\!\!
    \begin{bmatrix}
        -\Sc{x}_{\site{\R'}{(3,1)},\l} \\ \hphantom{-} \Sc{y}_{\site{\R'}{(3,1)},\l} \\ -\Sc{z}_{\site{\R'}{(3,1)},\l}
    \end{bmatrix}\!\!,
\end{equation}
while the spins in dimers 2 and 4 transform as
\begin{equation}
    \begin{bmatrix}
        \Sc{x}_{\site{\R}{(2,4)},\l} \\ \Sc{y}_{\site{\R}{(2,4)},\l} \\ \Sc{z}_{\site{\R}{(2,4)},\l}
    \end{bmatrix}
    \overset{M_x}{\rightarrow}
    \begin{bmatrix}
        \hphantom{-} \Sc{x}_{\site{\R'}{(4,2)},\l} \\ - \Sc{y}_{\site{\R'}{(4,2)},\l} \\ - \Sc{z}_{\site{\R'}{(4,2)},\l}
    \end{bmatrix},
\end{equation}
\end{subequations}
where $\R = i_x \hat{x} + i_y \hat{y}$ and $\R' = M_x(\R) = - i_x \hat{x} + i_y \hat{y}$ is the mirror image of $\R$ under $M_x$. The layer index $\l$ takes the values $T$ or $B$ for spins in the top or bottom layers, respectively.
In \app{app:symmetries}, we show that the composite transformation $\tilde{C}^z_2\circ M_x$ is a symmetry of the Hamiltonian. As an example, we focus on the transformation of a term with ferromagnetic interactions in \eqn{eqn:spin-ham-II}, such as
\begin{align}
H^{(x,y)}_\R := \K_1 \Scs{x}{y}_{\site{\R}{1},\l}\Scs{x}{y}_{\site{\R}{4},\l}-\K_1 \Scs{x}{y}_{\site{\R}{2},\l}\Scs{x}{y}_{\site{\R}{3},\l}
\end{align}
which transforms as 
\begin{equation}
\begin{split}
    H^{(x,y)}_\R &\overset{M_x}{\to}
    \K_1 \Scs{x}{y}_{\site{\R'}{3},\l}\Scs{x}{y}_{\site{\R'}{2},\l}-\K_1 \Scs{x}{y}_{\site{\R'}{4},\l} \Scs{x}{y}_{\site{\R'}{1},\l}
    \\
    &\overset{\tilde{C}^z_2}{\to}
    -\K_1 \Scs{x}{y}_{\site{\R'}{3},\l}\Scs{x}{y}_{\site{\R'}{2},\l}+\K_1 \Scs{x}{y}_{\site{\R'}{4},\l} \Scs{x}{y}_{\site{\R'}{1},\l}
    \\
    &=H^{(x,y)}_{\R'}.
\end{split}
\end{equation}
We see from above, since $H^{(x,y)}_\R \overset{\tilde{C}^z_2\circ M_x}{\to} H^{(x,y)}_{\R'}$, where $\R'$ is a primitive lattice vector, the sum $\sum_\R H^{(x,y)}_\R$ remains invariant under the transformation. Similarly, the sum of the rest of the Heisenberg interaction terms in \eqn{eqn:spin-ham-I} also remains invariant\ifdefined\useapp, see \app{app:symmetries} for further details\fi. This establishes $\tilde{C}^z_2\circ M_x$ as a one of the mirror-like symmetries for our first model. 

Searching for the second symmetry, we find that the usual-mirror reflection $M_y$, about a line parallel to the $x$-axis and bisecting a unit cell, leaves the spin Hamiltonian in \eqn{eqn:spin-ham-I} invariant. Under this symmetry operation, the spins transform as
\begin{subequations}
\begin{align}
    \begin{bmatrix}
        \Sc{x}_{\site{\R}{(2,3)},\l} \\ \Sc{y}_{\site{\R}{(2,3)},\l} \\ \Sc{z}_{\site{\R}{(2,3)},\l}
    \end{bmatrix}
    &\overset{M_y}{\rightarrow}
    \begin{bmatrix}
        - \Sc{x}_{\site{\R'}{(3,2)},\l} \\ \hphantom{-} \Sc{y}_{\site{\R'}{(3,2)},\l} \\ - \Sc{z}_{\site{\R'}{(3,2)},\l}
    \end{bmatrix}
    \\
    \begin{bmatrix}
        \Sc{x}_{\site{\R}{(1,4)},\l} \\ \Sc{y}_{\site{\R}{(1,4)},\l} \\ \Sc{z}_{\site{\R}{(1,4)},\l}
    \end{bmatrix}
    &\overset{M_y}{\rightarrow}
    \begin{bmatrix}
        - \Sc{x}_{\site{\R'}{(4,1)},\l} \\ \hphantom{-} \Sc{y}_{\site{\R'}{(4,1)},\l} \\ - \Sc{z}_{\site{\R'}{(4,1)},\l}
    \end{bmatrix},
\end{align}
\end{subequations}
where $\R' = M_y(\R) = i_x\hat{x} - i_y\hat{y}$.
\ifdefined\useapp 
We provide further details on the invariance of the terms in \eqn{eqn:spin-ham-I} under $M_y$ in \app{app:symmetries}.
\fi

The two symmetries $\tilde{C}^z_2\circ M_x$ and $M_y$ that we found above do not mix the triplon flavors, but admit different representations in the three decoupled triplon flavor sectors. As previously stated, we shall focus on the $t_x$ and $t_y$ sectors since these are the sectors that contribute to the HOT triplon phase. In the $t_x$ sector, the matrix representations $\mx^{(x)}$ and $\my^{(x)}$ for the symmetries that transform the bosonic Nambu \spinor\ $\Fi_{\R,x}$ in  \eqn{eqn:trip-ham-real-I} are
\begin{subequations}\label{eq:m_matrices_model1}
\begin{equation}
\begin{split}
    \tilde{C}^z_2\circ M_x &\equiv \mx^{(x)} = - \sigma_1 \eta_3 \tau_0
    \\
    M_y &\equiv \my^{(x)} = - \sigma_1 \eta_1 \tau_0,
\end{split}
\end{equation}
whereas for the $t_y$ sector, the matrix representations $\mx^{(y)}$ and $\my^{(y)}$ operate as
\begin{equation}
\begin{split}
    \tilde{C}^z_2\circ M_x &\equiv \mx^{(y)} = \sigma_1 \eta_3 \tau_0
    \\
    M_y &\equiv \my^{(y)} = \sigma_1 \eta_1 \tau_0
\end{split}
\end{equation}
\end{subequations}
on $\Fi_{\R,y}$. 
In both cases, the Bloch Hamiltonians for Model I in \eqns{eqn:trip-bloch-ham-model-I} and \ref{eqn:trip-bloch-ham-model-I-aniso} respectively satisfy $H_{\k}^\text{tBBH} = {\m{\alpha}^{(\beta)}}^\dagger H_{M_\alpha \k}^\text{tBBH} \m{\alpha}^{(\beta)}$, where $\alpha,\beta \in \{x,y\}$, leaving the full Hamiltonian invariant. 
Using the representations for the symmetry operations in \eqn{eq:m_matrices_model1}, we find $\big\{\mx^{(\alpha)},\my^{(\alpha)}\big\}=0$, where $\alpha \in \{x,y\}$, making the non-commuting property of the symmetries manifest, and hence fulfilling one of the main requirements for obtaining a HOT phase.

\subsection{Model II: \modII~model}
To discuss the symmetries of our second model, we define the DM and $\Gamma$ interactions terms from \eqn{eqn:spin-ham-II} at the lattice position ($\site{\R}{a}$) as 
\begin{equation}
\begin{split}
    D_{\R,a} &= \sum_{l}(-1)^a D\left(\Sc{y}_{\site{\R}{a},\l} \Sc{x}_{\site{\R+\hat{y}}{a},\l}-\Sc{x}_{\site{\R}{a},\l} \Sc{y}_{\site{\R+\hat{y}}{a},\l}\right),
    \\
    \Gamma_{\R,a} &= \sum_{l}(-1)^a \Gamma\left(\Sc{y}_{\site{\R}{a},\l} \Sc{x}_{\site{\R+\hat{y}}{a},\l}+\Sc{x}_{\site{\R}{a},\l} \Sc{y}_{\site{\R+\hat{y}}{a},\l}\right),
\end{split}
\end{equation}
respectively. A mirror reflection ($M_x$) about a line parallel to the $y$ axis and passing through the center of a unit cell transforms the spin operators as
\begin{align}
    \begin{bmatrix}
        \Sc{x}_{\site{\R}{(1,2)},\l} \\ \Sc{y}_{\site{\R}{(1,2)},\l} \\ \Sc{z}_{\site{\R}{(1,2)},\l}
    \end{bmatrix}
    \overset{M_x}{\longrightarrow}
    \begin{bmatrix}
        \hphantom{-} \Sc{x}_{\site{\R'}{(2,1)},\l} \\ - \Sc{y}_{\site{\R'}{(2,1)},\l} \\ - \Sc{z}_{\site{\R'}{(2,1)},\l}
    \end{bmatrix},
\end{align}
where $\R'=M_x(\R)$, or more explicitly $\R = i_x\hat{x} + i_y\hat{y}$ and $\R' = -i_x\hat{x} + i_y\hat{y}$.
Therefore, under $M_x$, the DM terms $D_{\R,a}$ transforms according to $D_{\R,1}\overset{M_x}{\rightarrow} D_{\R',2}$ and $D_{\R,2}\overset{M_x}{\rightarrow} D_{\R',1}$---that is, up to a primitive lattice translation, the DM term for dimer $1$ maps to the term for dimer $2$, and vice-versa. 
Since $\R'$ is a primitive lattice vector, the sum $\sum_{\R,a}D_{\R,a}$ over lattice vectors remains invariant; the same goes for the sum of Gamma interactions $\sum_{\R,a}\Gamma_{\R,a}$. The other terms in \eqn{eqn:spin-ham-II} can be shown to remain invariant as well\ifdefined\useapp~(see \app{app:symmetries} for details)\fi, implying that $M_x$ is indeed a symmetry of the Hamiltonian in \eqn{eqn:spin-ham-II}.

The second symmetry for the model is constructed by composing two operations: a mirror reflection $M_y$ about the $x$ axis passing between two unit cells followed by a $\pi/2$ global spin rotation $C^{z}_4$ about the $\hat{z}$ direction. The spin operators transform as
\begin{equation}
    \begin{bmatrix}
        \Sc{x}_{\site{\R}{a},\l} \\ \Sc{y}_{\site{\R}{a},\l} \\ \Sc{z}_{\site{\R}{a},\l}
    \end{bmatrix}
    \overset{M_y}{\rightarrow}
    \begin{bmatrix}
        - \Sc{x}_{\site{\R'}{a},\l} \\ \hphantom{-} \Sc{y}_{\site{\R'}{a},\l} \\ - \Sc{z}_{\site{\R'}{a},\l}
    \end{bmatrix}
    \overset{C^{z}_4}{\rightarrow}
    \begin{bmatrix}
        \hphantom{-} \Sc{y}_{\site{\R'}{a},\l} \\ \hphantom{-} \Sc{x}_{\site{\R'}{a},\l} \\ - \Sc{z}_{\site{\R'}{a},\l}
    \end{bmatrix},
\end{equation}
where, $\R' = M_y(\R) = i_x\hat{x} - i_y\hat{y}$. As a result, the DM and $\Gamma$ terms transform according to $D_{\R_1,a}\to D_{\R_2,a}$ and  $\Gamma_{\R_1,a}\to \Gamma_{\R_2,a}$, where $\R_1 = i_x\hat{x} + i_y\hat{y}$ and $\R_2 = i_x\hat{x} - (i_y \!+\! 1)\hat{y}$. Since $\R_2$ is a primitive lattice vector, the sums $\sum_{\R,a}D_{\R,a}$ and $\sum_{\R,a}\Gamma_{\R,a}$ are invariant under $C^{z}_4\circ M_y$. Similarly, the Heisenberg terms in \eqn{eqn:spin-ham-II} can be shown to remain invariant as well \ifdefined\useapp (see \app{app:symmetries})\fi, thereby establishing $C^z_4\circ M_y$ as another mirror-like symmetry for our second model.

The non-commuting nature of the two symmetries $M_x$ and $C^z_4\circ M_y$ can be verified by writing their representations in the subspace of triplons contributing to the HOT phase, i.e.\ the $x$ and $y$ triplons. 
In this subspace, apart from changing $\R\to M_\alpha(\R)$, the symmetries are represented by matrices
\begin{subequations}\label{eqn:symm-rep-trip-II}
\begin{align}
    M_x\equiv \mx &= \sigma_1 \eta_3 \tau_0
    \\
    C^z_4\circ M_y\equiv \my &= \sigma_0 \eta_1 \tau_0,
\end{align}
\end{subequations}
which act on the column vector $\Si_\R$ (see \eqn{eqn:trip-ham-real-II}), and constrain the Bloch Hamiltonian $H_{\k}^\text{tSOC}$ as $H_{\k}^\text{tSOC} = \m{\alpha}^\dagger H_{M_\alpha \k}^\text{tSOC} \m{\alpha}$, where $\alpha\in\{x,y\}$. The matrices $\mx$ and $\my$ satisfy $\{\mx,\my\}=0$ and hence, as expected, do not commute.

In the presence of a Zeeman field, one or both of the symmetries break depending on whether a staggered field $\hs$ (\eqn{eqn:hs-def}) or homogeneous field $h$ (\eqn{eqn:h-def}) is applied. For the staggered field case, $M_x$ (or $\mx$) is preserved, while $C^z_4\circ M_y$ ($\my$) is broken. The latter gets replaced by $\TR\circ C^z_4\circ M_y$, i.e., $C^z_4\circ M_y$ followed by a time-reversal operation $\TR$, as the new symmetry. When a homogeneous field is applied, the new symmetry $\TR\circ C^z_4\circ M_y$ also works as the replacement for the broken $C^z_4\circ M_y$ symmetry. However this time, in addition to $C^z_4\circ M_y$, the symmetry $M_x$ also breaks and is replaced by $\TR\circ M_x$. The matrix representations for the new symmetries in the relevant triplon sectors are obtained by left-multiplying the complex-conjugation operator $\cconj$ with the matrices $\mx$, $\my$ in \eqn{eqn:symm-rep-trip-II}. The modified representations still continue to anti-commute\ifdefined\useapp, see \app{app:symmetries} for details\fi. 
\section{Topological invariants}\label{sec:topo-inv}

It has been shown~\mycite{benalcazar2017quantized, benalcazar2017moments} that we can characterize the 1D edge of a 2D fermionic HOT phase using standard first-order topological invariants such as polarization~\mycite{Resta1998Position}. In turn, the bulk of the HOT phase can then be characterized using the invariants of their effective edge theories. 
The idea of using edge polarization as a topological invariant has been generalized to bosonic systems in the context of magnons~\cite{hirosawa2020magnonic, park2021hinge}. Here, we apply this framework to the triplon systems introduced in \Sec{sec:triplons}; however, we remark here that the term \emph{polarization} is used to refer to the topological invariant, and not to a physical polarization of the system. It has been demonstrated~\mycite{benalcazar2017quantized, benalcazar2017moments} that the topological nature of the edge polarizations can be characterized using Wannier Hamiltonians (a bulk property) without introducing physical edges in the model. 
We can obtain the Wannier Hamiltonians by constructing Wilson loop operators $\W$ that traverse the first Brillouin zone (BZ) once along either the $\alpha=x$ or $y$ directions starting from a base momentum $\k$. 
Defining the overlap matrix
\begin{equation}\label{eq:overlap-mat}
    \Fak = [\Fak]_{m,n} :=\bra{u_{m,\k+\dk}}\Sigma\ket{u_{n,\k}}
\end{equation}
for each momentum $\k$, the loop operator $\W$ is conveniently expressed as an ordered matrix product of the overlap matrices 
\begin{align}
    \WL^\alpha_\k=&F_{\alpha,\k+N_\alpha\dk}\cdots F_{\alpha,\k+\dk}F_{\alpha\k}
 \end{align}
following the loop.
Here, $\Sigma := 1_n\otimes\tau_3$ is the symplectic identity, $\ket{u_{n,\k}}$ are the eight-component particle-like Bloch vectors obtained by solving the Bloch Hamiltonians of \eqn{eqn:trip-bloch-ham-model-II} and \ref{eqn:trip-bloch-ham-model-I-aniso}, $N_{\alpha=x,y}$ are the number of unit cells along the $x$ or $y$ directions and $\dk=2\pi\hat{\alpha}/N_{\alpha}$. The band indices $m$ and $n$ label the rows and columns of the matrix $\Fak$ and run over a subset of bands in the system which we choose to be bands 1 and 2 for each our models (see \fig{fig:band-structures}(e), (f)). This choice of bands is guided by the analogy with the fermionic BBH model, for which bands 1 and 2 would be the filled bands. We mention here that, unlike in the fermion version, $\Fak$ is defined in terms of an inner product with matrix $\Sigma$~\footnote{The paraunitarity condition implies that the (particle-like) eigenstates at a given $\k$ are orthonormal with respect to this inner product with metric $\Sigma$.}; a similar definition has been used in the context of magnons---another bosonic BdG system---in Ref.~\onlinecite{park2021hinge}.

The Wannier Hamiltonians $\WH(\k)$, one for each direction $x$ and $y$, are defined by the loop operators via
\begin{equation}
    \exp\big(2 \pi \ci \, \WH(\k)\big) := \W.
\end{equation}
The Wannier Hamiltonian for the direction $\alpha$ is adiabatically connected (topologically equivalent) to the physical Hamiltonian for the edge perpendicular to $\alpha$~\mycite{FidkowskiPRL2011, benalcazar2017quantized}, allowing us to use it to characterize the edge theories. 
The eigenvalues of $\WH(\k)$ give the Wannier bands $\WB(\k)$ ($j=1,2$), which are defined modulo $1$, whereas the eigenvectors $\ket{\nu^{\alpha}_j(\k)}$ can be used to obtain~\mycite{FidkowskiPRL2011, benalcazar2017quantized} the Wannier band subspace spanned by the states $\WS$ in the original eight-component Nambu basis .
Here, we have used the same notation for denoting Wannier bands, subspaces, etc.\ as in Ref.~\onlinecite{benalcazar2017quantized}. The bands $\WB(\k)$, constructed from Wilson loops along the direction $\alpha$, depend only on the momentum component perpendicular to $\alpha$; e.g., $\WBc{x}(k_x,k_y)$ depends on $k_y$ only.  Since the matrices $\Fak$ in each of our models are constructed using Bloch bands 1 and 2, the number of Wannier bands at any given $\k$ is two. 

A useful feature of the Wannier bands is that they are gapped inside a HOT phase and become gapless when the underlying system transitions to a different topological phase. The gap closing of the Wannier bands is not necessarily accompanied with the closing of the bulk bandgap. As a result, the associated states $\WS$ can carry topological inavriants of their own. For the class of 2D HOT phases that we consider in this paper, this topological invariant was shown to be the polarization of the Wannier bands, or \emph{Wannier-sector polarization}~\mycite{benalcazar2017quantized}. To elaborate on this further, we focus on the Wannier Hamiltonian $\WHc{x}$ for the Wilson loops along the $x$ direction. If we define the overlap amplitude $\Fpol_{j,\k}=\bWSkc{\k+\dk}\Sigma\WS$ (where $\dk=(0,2\pi/N_y)$), we can construct yet another set of Wilson loops,
\begin{align}\label{eqn:nWL}
    \nWl_{y,k_x}=\Fpol_{j,\k+N_y\dk}\cdots \Fpol_{j,\k},
\end{align}
one for each Wannier band $j \in \{1,2\}$, that run along the $y$ direction. This time, however, the loops are formulated using the Wannier band basis $\WS$ instead of the Bloch states $\ket{u_{n,\k}}$ (see \eqn{eq:overlap-mat}) and the overlap amplitudes $\Fpol_{j,\k}$ are scalars instead of matrices. Following the convention in Ref.~\onlinecite{benalcazar2017quantized}, we call these new loops \emph{nested} Wilson loops. Since the $\Fpol_{j,\k}$ are scalars, the product $\nWl_{y,k_x}$ only depends on $k_x$. The polarization for the states $\WSck{x}$ can be written in terms of the nested Wilson loops as follows:
\begin{align}
    \pyj=-\frac{\ci}{2\pi N_x}\sum_{k_x}\log \nWl_{y,k_x}.
\end{align}
The Wannier-sector polarization $\pyj$, defined modulo 1, characterizes the topology on the $x$ edge. A value of $\pyj=0$ indicates that the edge has a trivial topology, while $\pyj=1/2$ implies a non-trivial edge topology. Similar to the fermionic BBH model, the quantization of $\pyj$ is a result of the constraints imposed on the polarization due to the mirror-like symmetries we introduced in \Sec{sec:symmetries}. The symmetries also constrain the polarizations of the two Wannier bands, $j \in \{1,2\}$, to be equal, i.e., $\pyjb{1}=\pyjb{2} \mod 1$. Therefore, we refer to the Wannier-sector polarizations $\pyjb{{1,2}}$ collectively as $p_y$. We have verified numerically that these constraints hold. In contrast to the fermionic case, our definition of the Wannier-sector polarization, given in \eqn{eqn:nWL}, uses the symplectic identity $\Sigma$ instead of the usual matrix identity. 
Similar definitions for computing polarizations of
Bloch bands~\mycite{engelhardt2015PRA} and of Wannier bands~\mycite{hirosawa2020magnonic} in bosonic BdG systems can be found in existing literature.

Following the same set of steps, we can define the Wannier-sector polarization $p_x$ starting from the Wannier Hamiltonian $\WHc{y}(\k)$ to characterize the $y$-edge topology. The polarization $p_x$ will also satisfy the same symmetry constraints as $p_y$, and will take values $0$ or $1/2$ modulo 1. Therefore, we can use the pair of Wannier-sector polarizations $\mbf{p}=(p_x,p_y)$ to further classify the HOT triplon phases that appear in the phase-diagrams (\fig{fig:phase-diagrams}) of our models.

\paragraph{Model I (\modI~model):} 
Numerically computing the Wannier-sector polarizations for the first model, we find the HOT-triplon phase in \fig{fig:phase-diagrams}(a), obtained when $\Kh_{1}<\Kh_{2}$ and $\Kv_{1}<\Kv_{2}$, has $\p=(1/2,1/2)$, implying that both the $y$-edge and $x$-edge polarizations are quantized to $1/2$. In the HOT phase, the Wannier bands $\WBjs(\k)$ for both the $\alpha \in \{x,y\}$ directions are gapped. We show the two gapped bands $\WBc{x}(\k)$, calculated from Wilson loops running along the $x$ direction, in  \fig{fig:phase-diagrams}(b) bottom inset. Transitions to other phases occur when either the $x$- or $y$-direction Wannier-band gap closes modulo 1. For example, when $\Kv_1\to\Kv_2$, the gap between the bands $\WBc{x}(\k)$, $j \in \{1,2\}$, begins to close, as seen from \fig{fig:phase-diagrams}(b) top inset. Hence, the line $\Kv_1=\Kv_2$ is a phase boundary (see dashed horizontal line in \fig{fig:phase-diagrams}(a)). Similarly, the line $\Kh_1=\Kh_2$ (represented by a dashed vertical line) forms another phase boundary, along which the gap between the Wannier bands $\WBc{y}(\k)(j=1,2)$ closes. The Wannier-sector polarizations of the resulting additional phases are $\p=(0,1/2)$ when $\Kv_1>\Kv_2$ and $\Kh_1<\Kh_2$,  $\p=(1/2,0)$ when $\Kv_1<\Kv_2$ and $\Kh_1>\Kh_2$, and $\p=(0,0)$ when both $\Khv_1>\Khv_2$. Therefore, the phases of \fig{fig:phase-diagrams}(a) are classified as $\mathbb{Z}_2\times \mathbb{Z}_2$, exactly like the fermion BBH model. The dependence of the minimum of the two Wannier-band gaps, $\min\{(\WBcj{x}{1}-\WBcj{x}{2})\mod 1,(\WBcj{y}{1}-\WBcj{y}{2})\mod 1\}$, on $\Kh_1, \Kv_{1}$ is shown as a density plot in the same figure.

\paragraph{Model II (\modII~model):}
For our second model, where the HOT phase of SOC triplons is destroyed by applying either a staggered or a homogeneous Zeeman field, we find two different topological classifications. The Wannier-sector polarizations $\p$ for the phases in \fig{fig:phase-diagrams}(c), obtained when a staggered Zeeman field $\hs$ is applied, are in one-to-one correspondence with the phase diagram (\fig{fig:phase-diagrams}(a)) for our first model. The only difference is that the role of $\Kv_1$ in the first model is now fulfilled by the staggered field strength $\hs$, and $\Kh_{1,2}$ are relabelled as $\K_{1,2}$, respectively; the line $\hs=\hcstag$ (\eqn{eqn:hcstag}) now marking one of the phase boundaries (see dashed-horizontal line in \fig{fig:phase-diagrams}(c)). Therefore, the topological phases for the staggered-field case are again classified as $\mathbb{Z}_2\times\mathbb{Z}_2$. Moving on to the phase diagram for the homogeneous field in \fig{fig:phase-diagrams}(d), we find that we can classify the phases using the Wannier-sector polarization $\p$ only when $h<\hchomI$. The HOT phase that occurs when $\K_1<\K_2$ and $h<\hchomI$ has polarizations $\p=(1/2,1/2)$. Like the previous cases, the transition to a phase with $\p=(0,1/2)$ takes place via a Wannier-band gap closing along the vertical line segment $\K_1=\K_2$ (dashed line \fig{fig:phase-diagrams}(d)). When $\hchomI\leq h\leq \hchomII$, the \emph{bulk} is gapless and therefore the system cannot host in-gap states. Beyond $h>\hchomII$, although the bulk becomes gapped once again, it is hard to determine from numerics whether the Wannier bands are truly gapped. 
Consequently, the numerical calculation of the Wannier sector polarizations also becomes challenging.
Nonetheless, in this region of parameter space, real-space numerics do not reveal any protected edge states or corner states. This \emph{may} indicate that the construction of the Wannier Hamiltonian using only two Bloch bands 1, 2 may no longer be adequate and must be generalized to include all four bands 1 to 4 to explore this region of the phase diagram; this will be explored in future work.

Before we end this section, we mention that the Wannier-sector polarizations can be calculated without involving the $\Sigma$ matrix if we work in the limit of large dimer strength $\J$. In this limit, cross amplitudes between the particle-like and hole-like sectors in the BdG Hamiltonians (\eqns{eqn:trip-bloch-ham-model-I-aniso}, \ref{eqn:trip-bloch-ham-model-II})
get suppressed as $\mathcal{O}(1/\J)$ and we can work with the Hamiltonian projected on to the particle sector (such a projection is discussed in \Sec{subsec:tSOC-ET} and \app{subsec:edgehamiltonians-modelII-effedgetheories}). The projected Hamiltonian is Hermitian and can be diagonalized using unitary matrices instead of paraunitary ones. Since the non-trivial topology arises in the excited triplon sector of our models, the projected Hamiltonian, for each model, is topologically equivalent and adiabatically connected to the original unprojected Hamiltonian. Hence, both the Hamiltonians share the same topological invariants for the triplon sector.
The eigenstates of the projected Hamiltonian are orthonormal with respect to the standard inner product, and therefore the original fermionic definitions for Wilson loops, etc., can be used to calculate the Wannier-sector polarizations. We have numerically verified that this gives the same answer as the previous analysis.
 
\section{Edge theory and Shared corner modes}\label{sec:edge-theo}

The corner modes of a HOT phase show additional features compared to edge modes at the junction of a 1D trivial and a first-order topological system. In particular, the HOT-corner modes are \emph{shared} by the topological edge theories that intersect at the corner. This shared nature of a corner mode can be verified by finding the effective continuum theories of the intersecting edges, e.g., right ($x=0$) and top ($y=0$) edges (see \fig{fig:models}(b),(d)), and then introducing corners into the resulting 1D theories. Hence, the corner can be considered as the edge of a 1D edge theory. Since the edge theory of a 2D HOT phase is described by first-order topology, the corner mode can be recovered following a Jackiw-Rebbi-type~\mycite{JackiwPRD1976, benalcazar2017quantized} approach and tuning a mass-like term across \emph{zero} at some point along the spatial direction. We now describe finding a shared corner mode for each of our two models in more detail. 

\subsection{Model I: \modI \ model}\label{subsec:tBBH-ET}
We work with the anisotropic version of the model again (see \eqn{eqn:trip-bloch-ham-model-I-aniso}) in which the couplings $\Kh_{1,2}$ along the $\hat{x}$ direction are in general distinct from the couplings $\Kv_{1,2}$ along $\hat{y}$. Doing so allows us to isolate the edge Hamiltonian for the $x=0$ edge from the $y=0$ edge and vice-versa. We obtain the continuum Hamiltonian,
\begin{multline}\label{eqn:HLE-model-I}
    \HLEI = \J \sigma_0\eta_0\tau_3 + \Bigl( m_x \sigma_1\eta_0 + v_x (-\ci\partial_x) \sigma_2\eta_3
    \\
    - m_y \sigma_2\eta_2 + v_y (-\ci\partial_y) \sigma_2\eta_1 \Bigr) \! (\tau_3 \!-\! \ci\tau_2),
\end{multline}
by expanding the Bloch Hamiltonian in \eqn{eqn:trip-bloch-ham-model-I-aniso} around the bandgap-closing momentum $\kgap$, which for positive 
$\Khv_{1,2}$
is $\kgap=(\pi,\pi)$ (see \app{app:bandstructures}), and then taking the continuum limit by transforming $q_x \rightarrow -\ci \partial_x$ and $q_y \rightarrow -\ci \partial_y$, where $\mbf{q}=(q_x,q_y)$ is a small deviation from $\kgap$. The mass-like terms $m_x$ and $m_y$ in \eqn{eqn:HLE-model-I} are related to the original couplings as $m_x=(\Kh_1 - \Kh_2)/2$ and $m_y=(\Kv_1 - \Kv_2)/2$, while the coefficients $v_x$ and $v_y$ are given by $v_x = \Kh_2/2$, $v_y = \Kv_2/2$. Assuming positive $v_x$ and $v_y$, when the mass term $m_x (m_y) < 0$, the phase on the $y=0$ ($x=0$) edge is topological, while $m_x (m_y)>0$ makes the phase trivial. Setting $m_x (m_y) =0$ produces a gapless phase on the $y=0$ ($x=0$) edge. Therefore, we obtain the Hamiltonian of the $y=0$ (top) edge,
\begin{align}\label{eqn:HLEItop}
  \HLEItop &= \J \sigma_0\tau_3 + \bigl( m_x \sigma_1 + v_x (-\ci\partial_x) \sigma_2 \bigr) (\tau_3 - \ci\tau_2),
\end{align}
by introducing a domain wall $m_y(x,y)=\tanh(y)$ along the top edge, and solving for modes that disperse along the $x$ direction but decay exponentially in the $y$ direction, see \app{app:edge_hamiltonians} for details. The exact function chosen to model domain walls like $m_y(x,y)$ is not critical to our analysis; any function $f(y)$ interpolating between $m_y<0$ (topological) and $m_y>0$ (trivial) with a \emph{single} zero crossing is a valid choice. The corner mode for $\HLEItop$ can now be obtained by inserting yet another domain wall $m_x(x)=\tanh(x)$ and looking for a \emph{single} trapped mode. This process (see \app{app:edge_hamiltonians}) gives
\begin{align}
    \psitr=\begin{bmatrix} 1 \\ 0 \end{bmatrix}^{(\sigma)}
        \otimes
        \begin{bmatrix} 1 \\ 0 \end{bmatrix}^{(\eta)}
        \otimes
        \begin{bmatrix} 1 \\ 0 \end{bmatrix}^{(\tau)}
\end{align} 
as the corner mode living on the top-right corner that has the near-mid-bandgap energy $E=\J$. To find the corner mode supported by the right ($x=0$) edge, we reverse the order in which we introduce the domain walls. To be more specific, we first get the continuum Hamiltonian $\HLEIright$ for the right edge by inserting the domain wall $m_x(x,y)=\tanh(x)$, this time running along the right edge. We then introduce the second domain wall $m_y(y)=\tanh(y)$ in the 1D Hamiltonian $\HLEIright$ and find, yet again, a single trapped mode $\psirt$ at the near-mid-bandgap energy $E=\J$\ifdefined\useapp, see \app{app:edge_hamiltonians} for the explicit calculation\fi. Comparing $\psitr$ with $\psirt$ we find that they are \emph{identical}, thus confirming that the corner mode is shared between the top and right edges.

\subsection{Model II: \modII \ model}\label{subsec:tSOC-ET}

We now turn to the corner mode in the \modII\ model: we wish to show that it too is shared between the edge theories.
As suggested by the similarity between the phase diagram of the \modII\ model in the presence of a staggered field (\fig{fig:phase-diagrams}(b)) and that of the \modI\ model (\fig{fig:phase-diagrams}(a)), the staggered-field strength $\hs$ can serve as a tuning parameter for introducing edges into the \modII\ model---specifically, $\hs$ plays a role analogous to that of $\Kv_1$ in the \modI~model. With this in mind, we develop an analysis similar to the one in \Sec{subsec:tBBH-ET}, starting from the Bloch Hamiltonian
\begin{equation}\label{eq:HtSOChs}
\Hhs = H_{\k}^\text{tSOC} 
- \hs \sigma_3 \eta_2 \tau_3,
\end{equation}
where $H_{\k}^\text{tSOC}$ was defined in \eqn{eqn:trip-bloch-ham-model-II}.
The band structure for $\Hhs$ in the HOT phase typically looks like \fig{fig:corner-modes}(f) even in the presence of a staggered field,  i.e., bands 1 and 2 are degenerate and separated by a bulk bandgap (for $D,\Gamma\neq 0$) from bands 3 and 4 which are also degenerate. The bulk bandgap closes at the gap-closing momentum $\kgap = (\pi,\pi/2)$ when $\K_1=\K_2$ and $\hs=\hcstag$ (see \eqn{eqn:hcstag} as well as \app{app:bandstructures}), and critical staggered field $\hcstag\approx D$ when  $|D| \ll \J$. The continuum Hamiltonian in the neighborhood of $\kgap$ is
\begin{multline}\label{eqn:HLEII}
    \HLEII = 
    \J\sigma_0\eta_0\tau_3 
\\
    + \Biggl( \!\! \frac{\K_1 \!\! - \!\! \K_2}{2}\sigma_1 \! - \! \frac{\K_2}{2} (-\ci\partial_x) \sigma_2 \!\! \Biggr) \eta_0 \left(\tau_3 \!-\! \ci\tau_2\right)
    \\
    - \sigma_3\eta_2\bigl( \hs \tau_0 - D\left(\tau_3 \!-\! \ci\tau_2\right) \bigr) 
    \\
    - \Gamma (-\ci\partial_y) \sigma_3\eta_1 \left(\tau_3\!-\!\ci\tau_2\right),
\end{multline}
where we have assumed $\K=0$ for simplicity. We have checked that the outcome of the calculation does not depend on $\K$ provided $|\K| < |\Gamma|$; see \app{app:edge_hamiltonians} for details. 

Identifying mass terms in \eqn{eqn:HLEII} analogous to the ones in \eqn{eqn:HLE-model-I} is straightforward if we work in the large-$\J$ limit. This is justified because the dominant energy scale $\J$ sets the size of the gap between the singlet ground state and triplon excitations, meaning the model is asymptotically connected to the $\J\to \infty$ limit. Consequently, the finite-$\J$ model is topologically equivalent to the leading-order Hamiltonian in the large-$\J$ limit.
The latter Hamiltonian is obtained by projecting the finite-$\J$ Hamiltonian onto the particle-like sector of the BdG space---in effect performing a first-order Schrieffer-Wolff transformation (Refs.~\onlinecite{massarelli_arxiv_2021,bravyi2011schrieffer,winkler2003quasi}, also see \app{app:edge_hamiltonians}) onto the positive eigenstates of $\J\sigma_0\eta_0\tau_3$. The effective (Hermitian) Hamiltonian describing the positive-energy bands is then 
\begin{multline}
    \HLEIIeff = \J\sigma_0\eta_0 + \bigl( m_x\sigma_1 - v_x (-\ci\partial_x) \sigma_2 \bigr) \eta_0
    \\
    - \sigma_3 \bigl( m_y\eta_2 + v_y (-\ci\partial_y) \eta_1 \bigr),
\end{multline}
where $m_x = (\K_1-\K_2)/2$, $m_y = \hs - D$, $v_x = \K_2/2$, and $v_y = \Gamma$. Since the critical field $\hcstag$ approaches $D$ in the large-$\J$ limit, the mass term $m_y$ indeed marks the topological transition that for finite $\J$ happens at $\hs=\hcstag$, represented by the horizontal dashed line in \fig{fig:phase-diagrams}(c).

We find the top-edge Hamiltonian $\HLEIIefftop$ following the same steps as for the \modI\ model: we first introduce a domain wall $m_y(x,y) = \tanh y$ along the top edge at $y=0$, and then write down the effective Hamiltonian 
\begin{equation}
    \HLEIIefftop = \J \sigma_0 + m_x \sigma_1 - v_x (-\ci\partial_x) \sigma_2
\end{equation}
for states that are exponentially confined to this top edge. As we did for the \modI \ model, we get to the corner mode for the current model by introducing another domain wall $m_x(x) = \tanh x$, leading to the mode
\begin{equation}
    \psitr = \begin{bmatrix} 0 \\ 1 \end{bmatrix}^{(\sigma)}
        \otimes \begin{bmatrix} 0 \\ 1 \end{bmatrix}^{(\eta)}
        \otimes \begin{bmatrix} 1 \\ 0 \end{bmatrix}^{(\tau)}
\end{equation}
localized at the top-right corner and having the midgap energy $E = \J$. 

Proceeding in reverse order, i.e., starting by introducing a right edge in the form of $m_x(x,y) = \tanh x$, leads to the 1D edge Hamiltonian $\HLEIIeffright$. An edge (or corner) is then introduced in this edge Hamiltonian by setting $m_y(y) = \tanh y$, leading to the corner mode $\psirt = \psitr$. Once again, this shows the corner mode is shared between the top and right edge theories. 
\section{Summary and outlook}\label{sec:conclusion}
We have shown in this paper that dimerized quantum paramagnets can host triplon modes which exhibit higher order topology and corner states at mid-bandgap energies. The 2D models proposed and studied here are spin variants of previously explored fermionic Hamiltonians or obtained by a staggered stacking of 1D dimer Hamiltonians which have been shown to exhibit first-order topology. We have provided numerical support for corner modes and the phase diagram of such HOT triplon phases and their transitions. In addition, we have provided analytical results on their symmetry protection and edge theories, and constructed their topological invariants. Our theoretical work thus opens up a new direction in the study of dimerized quantum magnets.

The wealth of quantum dimer magnets which have been explored over the years~\mycite{shastry1981, Smith1989, sasago1997, tanaka2001, sakurai2002, ruegg2003, ruegg2005, sebastian2006, giamarchi2008, ronnow2017, rossPRL2019} makes it is very likely that model Hamiltonians of the type we have studied can be realized in experiments.
Identifying specific materials candidates is a natural next step worth investigating. 
A possible guiding principle to narrow down the search for suitable candidate systems would be to look for materials with noncommuting mirror symmetries like those discussed in \Sec{sec:symmetries}, rather than usual mirror symmetries.

Another important open question, not substantially addressed in previous research on HOT phases in magnetic systems, is one of detecting the near-midgap energy corner modes. One possible route to probing such localized nonzero-energy spin excitations might be to excite them selectively via terahertz radiation~\mycite{armitagePRX2018,armitagePRL2020}, which can access typical magnetic exchange energy scales, and look for mid-bandgap energy terahertz absorption that scales with the number of corners in the sample. Another approach would be to look at site resolved spin-spin correlations using spin-Hall noise spectroscopy \mycite{Joshi2018Detecting}.
Further, it might be possible to employ recent advances in atomically resolved electron spin resonance measurements to probe the spatial localization of the excitations~\mycite{willkeScience2018}, or recently developed scanning techniques~\mycite{yacobyScience2017,yacoby2018,demler2018,demler2019} that can use nitrogen-vacancy centers in diamond to measure local magnetic moments and extract dynamical information via spin noise. In this context, the rare earth dimer systems \cite{rossPRL2019} could prove useful owing to the reduced exchange energy scales in these systems, which makes them amenable to such experimental techniques.
 
\begin{acknowledgments}
We acknowledge funding from the Natural Sciences and Engineering Research Council (NSERC) of Canada.
We gratefully recognize the computational resources provided to us by Compute Canada; in particular on the Niagara, Graham, and Cedar systems.
G.\ M.\ acknowledges funding from \emph{le Fonds de recherche du Qu\'ebec -- Nature et technologies}.
\end{acknowledgments}

\def\priorityI{}

\ifdefined\useapp
\appendix

\ifdefined\priorityI
    
\section{Band structures}\label{app:bandstructures}
In this section, we elaborate on the main features of the triplon band dispersions for our two models that we refer to in the main text.

\subsection{Model I: \modI~model}

The (doubly degenerate) eigenvalues of $\tau_3 H_{\k}^\text{tBBH}$ (from \eqn{eqn:trip-bloch-ham-model-I-aniso}) are $-E_{\k,+}$, $-E_{\k,+}$, $-E_{\k,-}$, $-E_{\k,-}$, $E_{\k,-}$, $E_{\k,-}$, $E_{\k,+}$, and $E_{\k,+}$, where 
\begin{multline}
  E_{\k,\pm}^2 = \J^2 \pm \J \Big(
    {\Kh_1}^2 + 2\Kh_1\Kh_2\cos k_x + {\Kh_2}^2
    \\
     + {\Kv_1}^2 + 2\Kv_1\Kv_2\cos k_y + {\Kv_2}^2
  \Big)^{1/2}.
\end{multline}
The minimum of $E_{\k,+}$ and maximum of $E_{\k,-}$ are respectively
\begin{equation}
\begin{split}
    E_+^\text{(min)} & = \sqrt{
    \J^2 + \J \sqrt{ \bigl(|\Kh_1| - |\Kh_2|\bigr)^2 + \bigl(|\Kv_1| - |\Kv_2|\bigr)^2 }
    }
    \\
    E_-^\text{(max)} &= \sqrt{
    \J^2 - \J \sqrt{ \bigl(|\Kh_1|-|\Kh_2|\bigr)^2 + \bigl(|\Kv_1|-|\Kv_2|\bigr)^2 },
    }
\end{split}
\end{equation}
showing that the bulk bandgap between $E_-$ and $E_+$ given in \eqn{eq:modIbulkgap} closes iff $|\Kh_1| = |\Kh_2|$ \emph{and} $|\Kv_1| = |\Kv_2|$. 
Furthermore, the closing occurs at momentum $\kgap = (k_{0x}, k_{0y})$, where
\begin{subequations}
\begin{align}
    k_{0x} &= 
    \begin{cases}
        \pi & \mbox{if } \Kh_1=\Kh_2
        \\
        0 & \mbox{if } \Kh_1=-\Kh_2
    \end{cases},
    \\
    k_{0y} &= 
    \begin{cases}
        \pi & \mbox{if } \Kv_1=\Kv_2
        \\
        0 & \mbox{if } \Kv_1=-\Kv_2
    \end{cases}.
\end{align}
\end{subequations}

\subsection{Model II: \modII~model}

In the absence of a Zeeman field, the (doubly degenerate) eigenvalues of $\tau_3 H_{\k}^\text{tSOC}$ are $-E_{\k,+}$, $-E_{\k,+}$, $-E_{\k,-}$, $-E_{\k,-}$, $E_{\k,-}$, $E_{\k,-}$, $E_{\k,+}$, and $E_{\k,+}$, where 
\begin{multline}
  E_{\k,\pm}^2 = \J^2 + 2\J \K\cos k_y
    \\
    \pm \J \biggl(
    2D^2\bigl(1-\cos(2k_y)\bigr) + 2\Gamma^2\bigl(1+\cos(2k_y)\bigr)
    \\
    + \K_1^2 + 2\K_1\K_2\cos k_y + \K_2^2
  \biggr)^{1/2}.
\end{multline}
There exists a full bandgap betwen $E_{\k,-}$ and $E_{\k,+}$ unless $|\K_1| = |\K_2|$ \emph{and} at least one of $D$ and $\Gamma$ is zero. 
In our work, we assume $D$ and $\Gamma$ are both nonzero, keeping the bands gapped at zero field even if $\K_1 = \K_2$. Furthermore, nonzero $D$ and $\Gamma$ is necessary to realize the HOT phase in this model. This is consistent with the findings in Ref.~\onlinecite{joshi2017topological}, albeit in the context of first-order topology.

In the presence of a staggered Zeeman field (see \eqn{eqn:hs-def}), the triplon Bloch Hamiltonian is given by
\begin{subequations}
\begin{equation}
\Hhs = H_{\k}^\text{tSOC} 
- \hs \sigma_3 \eta_2 \tau_3,
\end{equation}
while for a uniform field configuration (see \eqn{eqn:h-def}), we have
\begin{equation}\label{eqn:app_hh_def}
\Hh = H_{\k}^\text{tSOC} 
- h \sigma_0 \eta_2 \tau_3,
\end{equation}
\end{subequations}
where $H_{\k}^\text{tSOC}$ was defined in \eqn{eqn:trip-bloch-ham-model-II}. 
The Hamiltonian $\Hhs$ for the staggered field was already reported in \eqn{eq:HtSOChs}.

For both field configurations, the model does not admit a closed-form dispersion relation except at certain momenta. We now discuss the details of these band dispersions and critical field strengths in the presence of staggered and homogeneous Zeeman fields individually.

\subsubsection{Staggered Zeeman field}
In the presence of a staggered Zeeman field $\hs$, numerical investigations show that the eight bands of $\tau_3 H_{\hs}^\text{tSOC}(\k)$ (see \eqn{eq:HtSOChs}) remain doubly degenerate. 
The investigations also show that band touchings occur iff $|\K_1| = |\K_2|$ \emph{and} $|\hs| = |\hcstag|$, where $\hcstag := \bigl(\sqrt{\J^2+2\J D} - \sqrt{\J^2 - 2\J D}\bigr)/2 \approx D$ for $|D| \ll \J$. 
The band touchings occur at $\kgap = (k_{0x}, k_{0y})$, where 
\begin{subequations}
\begin{align}
    k_{0x} &= 
    \begin{cases}
        \pi & \mbox{if } \K_1=\hphantom{-}\K_2
        \\
        0 & \mbox{if } \K_1=-\K_2
    \end{cases},
    \\
    k_{0y} &= 
    \begin{cases}
        \hphantom{-}\frac{\pi}{2} & \mbox{if } \hs = \hphantom{-} \hcstag
        \\
        -\frac{\pi}{2} & \mbox{if } \hs = - \hcstag
    \end{cases}.
\end{align}
\end{subequations}
Analytical values for the crossings can be found from the closed-form expressions for the eigenvalues of $\tau_3 H_{\hs}^\text{tSOC}(\k)$ for $|\K_1| = |\K_2|$ at the above momenta.

\subsubsection{Homogeneous Zeeman field}
In the presence of a homogeneous Zeeman field $h$, the twofold degeneracy of the bands is broken. 
Numerical diagonalization of \eqn{eqn:app_hh_def} shows that band touchings occur only along the $k_y = \pm \pi/2$ lines. 
In terms of $f(k_x) := \sqrt{4D^2 + \K_1^2 + 2\K_1\K_2\cos k_x + \K_2^2}$, the (particle-like) eigenvalues along $k_y = \pm \pi/2$ are 
\begin{subequations}
\begin{align}
    E_1(k_x, \pm \pi/2) &= \sqrt{\J^2 - \J f(k_x)} - h
    \\
    E_2(k_x, \pm \pi/2) &= \sqrt{\J^2 - \J f(k_x)} + h
    \\
    E_3(k_x, \pm \pi/2) &= \sqrt{\J^2 + \J f(k_x)} - h
    \\
    E_4(k_x, \pm \pi/2) &= \sqrt{\J^2 + \J f(k_x)} + h.
\end{align}
\end{subequations}
With the degeneracy lifted, the near-midgap states now occur within the bandgap between $E_2$ and $E_3$.

Note that the maximum and minimum values reached by $f(k_x)$ are respectively $f_+$ and $f_-$, where
\begin{subequations}
\begin{align}
    f_+ &= \sqrt{ 4D^2 + \bigl( |\K_1| + |\K_2| \bigr)^2 }
    \\
    f_- &= \sqrt{ 4D^2 + \bigl( |\K_1| - |\K_2| \bigr)^2 }.
\end{align}
\end{subequations}
Using $f_\pm$ we define the two critical field values:
\begin{subequations}
\begin{align}
    \hchomI &:= \frac{ \sqrt{\J^2 + \J f_-} - \sqrt{\J^2 - \J f_-} }{2},
    \\
    \hchomII &:= \frac{ \sqrt{\J^2 + \J f_+} - \sqrt{\J^2 - \J f_+} }{2}.
\end{align}
\end{subequations}
It is now straightforward to see that there is a bandgap between $E_2$ and $E_3$ for $|h| < \hchomI$ and again for $|h| > \hchomII$. 
In the remaining regions (i.e.\ $\hchomI \leq |h| \leq \hchomII$), the middle bands touch with linear crossing points, except for the critical values themselves, at which the crossings are linear in one momentum direction and quadratic in the other.

\section{Derivation of effective edge Hamiltonians for triplons} \label{app:edge_hamiltonians}
In this appendix, we provide more details on the derivation of the effective edge theories and shared corner modes presented in \Sec{sec:edge-theo}. 
For each model, the effective edge Hamiltonians are obtained from the continuum theories taken about their respective bulk-bandgap-closing momenta. 
We discuss these details for the two models individually.

\begin{figure}
    \centering
\includegraphics[scale=1]{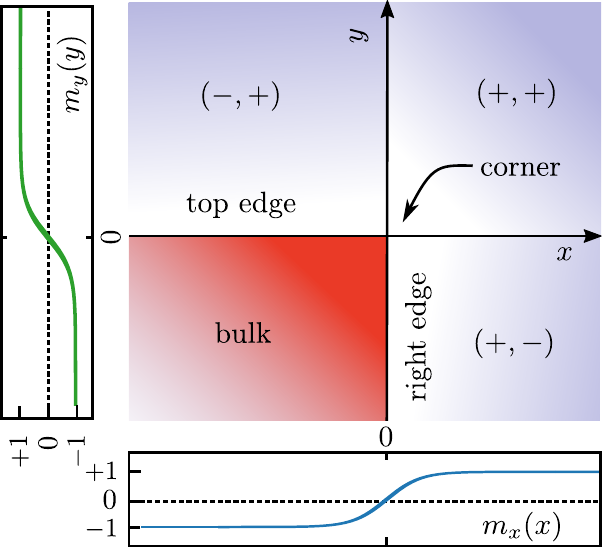}
\caption{Schematic representation of the mass terms $m_x$, $m_y$, as functions of position, that are used to isolate the corner mode shared by the top- and right-edge Hamiltonians.}
  \label{fig:mass_terms}
\end{figure}

\subsection{Model I: \modI~model}
\subsubsection{Effective edge theories}

Starting from the continuum Hamiltonian about $\k_0 = (\pi,\pi)$ (\eqn{eqn:HLE-model-I}) we isolate the ``top'' edge by introducing a domain wall $m_y(y)$ (see \fig{fig:mass_terms}) and seek edge states along the wall using a Jackiw-Rebbi ansatz,
\begin{equation} \label{eq:JB-modI-ansatz}
    \psi_{q_x}(x,y) = \re^{ - \gammatop \int_0^y \rd \tilde{y} \, m_y(\tilde{y})} \re^{ \ci x q_x} \psitop(q_x),
\end{equation}
where $q_x$ is the momentum along the edge and $\psitop(q_x)$ is an eight-component vector independent of $m_y(y)$, to be determined below. In the main text, we chose $m_y(y) = \tanh y$ for simplicity; however, the choice of $m_y$ is not unique. 
In principle, any smooth function $m_y(y)$ interpolating between $m_y < 0$ (topological) and $m_y > 0$ (trivial) with a \emph{single} zero crossing is an equally valid choice for a domain wall. Since $\psi_{q_x}(x,y)$ is meant to be an eigenvector of $\HLEI$, it must satisfy
\begin{equation}\label{eq:JB-SE}
    \HLEI \psi_{q_x}(x,y) = E_{q_x} \psi_{q_x}(x,y),
\end{equation}
where the eigenvalue $E_{q_x}$ gives the dispersion for the edge modes. If \eqn{eq:JB-SE} is to have a non-trivial solution such that $\psitop(q_x)$ is unique regardless of the choice of $m_y(y)$, $\psitop(q_x)$ must be an element of $\ker \bigl( \sigma_2 (\eta_2 - \ci \gammatop v_y \eta_1) (\tau_3-\ci\tau_2) \bigr)$. Nontrivial (normalizable) solutions of this sort exist only for $\gammatop = 1/v_y$ (see \eqn{eqn:HLE-model-I} for $v_y$), and are of the form
\begin{equation}\label{eq:tBBH-top-solutions}
    \psitop(q_x) = \chi^{(\sigma)}_{q_x} 
    \otimes 
    \begin{bmatrix} 1 \\ 0 \end{bmatrix}^{(\eta)}
    \otimes 
    \chi^{(\tau)}_{q_x},
\end{equation}
where $\chi^{(\sigma)}_{q_x}$ and $\chi^{(\tau)}_{q_x}$ are arbitrary two-component vectors.  With this choice, the effective Hamiltonian for the top edge is
\begin{equation}
    \HLEItop = \J \sigma_0\tau_3 + \left( m_x \sigma_1 + v_x q_x \sigma_2 \right) (\tau_3-\ci\tau_2),
\end{equation}
where $v_x$ was defined below \eqn{eqn:HLE-model-I}. We reported the above Hamiltonian for the top edge in \eqn{eqn:HLEItop} of the main text.

Alternatively, we can introduce a ``right'' edge with a domain wall $m_x(x)$ (see \fig{fig:mass_terms}). Using a Jackiw-Rebbi ansatz
\begin{equation}
    \psi_{q_y}(x,y) = \re^{ - \gammaright \int_0^x \rd \tilde{x} \, m_x(\tilde{x})} \re^{ \ci y q_y} \psiright(q_y),
\end{equation}
we seek eigenvectors of $\HLEI$:
\begin{equation}
    \HLEI \psi_{q_y}(x,y) = E_{q_y} \psi_{q_y}(x,y),
\end{equation}
where the eigenvalues $E_{q_y}$ give the dispersion for the top-edge modes. Again, for the ansatz to yield a solution regardless of the details of $m_x(x)$, $\psiright(q_y)$ must be part of $\ker \bigl( (\sigma_1 \eta_0 + \ci \gammaright v_x \sigma_2 \eta_3) (\tau_3-\ci\tau_2) \bigr)$. Solutions arise only if $\gammaright = 1 / v_x$. Though the space of solutions is of the same dimension as in \eqn{eq:tBBH-top-solutions}, $\psiright(q_y)$ does not factor in the same way. Instead, solutions are of the form
\begin{subequations}
\begin{equation}
    \psiright(q_y) = 
    \bigl( c_1(q_y) v_1 + c_2(q_y) v_2 \bigr)
    \otimes 
    \chi^{(\tau)}_{q_y},
\end{equation}
where $c_1(q_y)$ and $c_2(q_y)$ are arbitrary $q_y$-dependent coefficients, $\chi^{(\tau)}_{q_y}$ is an arbitrary two-component vector and
\begin{align}
    v_1 &:= \begin{bmatrix} 1 \\ 0 \end{bmatrix}^{(\sigma)}
    \otimes 
    \begin{bmatrix} 1 \\ 0 \end{bmatrix}^{(\eta)},
    &
    v_2 &:= \begin{bmatrix} 0 \\ 1 \end{bmatrix}^{(\sigma)}
    \otimes 
    \begin{bmatrix} 0 \\ 1 \end{bmatrix}^{(\eta)}.
\end{align}
\end{subequations}
The effective Hamiltonian for the right edge, now written in the basis of $(v_1, v_2)$ with Pauli matrices $\tilde{\sigma}_\alpha$, is
\begin{equation}\label{eq:app_rightedge_hamiltonian}
    \HLEIright = \J \tilde{\sigma}_0 \tau_3 + ( m_y \tilde{\sigma}_1 + v_y q_y \tilde{\sigma}_2 ) (\tau_3-\ci\tau_2).
\end{equation}

\subsubsection{Shared corner modes}

The corner mode for the top edge can now be obtained by inserting yet another domain wall in $m_x(x)$ at $x=0$ in the Hamiltonian $\HLEItop$ and looking for a \emph{single} trapped mode. Specifically, we look for eigenstates of the form
\begin{subequations}
    \begin{gather}
        \psitop(x) = \re^{ - \gammatopright \int_0^x \rd \tilde{x} \, m_x(\tilde{x})} \psitr,
        \\
        \psitr = \chi^{(\sigma)} \otimes 
        \begin{bmatrix} 1 \\ 0 \end{bmatrix}^{(\eta)}
        \otimes \chi^{(\tau)}.
    \end{gather}
\end{subequations}
For the solution to exist regardless of the details of $m_x(x)$, it should be part of $\ker (\sigma_1 + \ci \gammatopright v_x \sigma_2) (\tau_3 - \ci\tau_2)$. Furthermore, nontrivial solutions are of the form
\begin{equation}
    \psitr = \begin{bmatrix} 1 \\ 0 \end{bmatrix}^{(\sigma)}
        \otimes
        \begin{bmatrix} 1 \\ 0 \end{bmatrix}^{(\eta)}
        \otimes
        \chi^{(\tau)}
\end{equation}
and exist only if $\gammatopright = 1/v_x$. 
Two such solutions are obtained, and are related by particle-hole conjugation. The positive-energy solution is
\begin{equation}
    \psitr = \begin{bmatrix} 1 \\ 0 \end{bmatrix}^{(\sigma)}
        \otimes
        \begin{bmatrix} 1 \\ 0 \end{bmatrix}^{(\eta)}
        \otimes
        \begin{bmatrix} 1 \\ 0 \end{bmatrix}^{(\tau)}
\end{equation}
and appears at the near-midgap energy $E=\J$

On the other hand, the corner mode for the right edge is found by introducing a domain wall $m_y(y)$ in the Hamiltonian $\HLEIright$.
We use the ansatz
\begin{subequations}
    \begin{gather}
        \psitilderight(y) = \re^{ - \gammarighttop \int_0^y \rd \tilde{y} \, m_y(\tilde{y})} \psitildert,
        \\
        \psitildert = 
        \chi^{(\tilde{\eta})} \otimes \chi^{(\tau)},
    \end{gather}
\end{subequations}
where the tildes on the state vectors indicate that they are expressed in the basis of $(v_1, v_2)$ used in \eqn{eq:app_rightedge_hamiltonian}. A robust solution must be part of $\ker (\tilde{\sigma}_1 + \ci \gammarighttop v_y \tilde{\sigma}_2) (\tau_3 + \ci\tau_2)$, and only exists if $\gammarighttop = 1 / v_y$. The positive-energy solution, at $E = \J$, is
\begin{equation}
    \psitildert = \begin{bmatrix} 1 \\ 0 \end{bmatrix}^{(\tilde{\sigma})}
        \otimes
        \begin{bmatrix} 1 \\ 0 \end{bmatrix}^{(\tau)},
\end{equation}
or in the original representation,
\begin{equation}
    \psirt = \begin{bmatrix} 1 \\ 0 \end{bmatrix}^{(\sigma)}
        \otimes
        \begin{bmatrix} 1 \\ 0 \end{bmatrix}^{(\eta)}
        \otimes
        \begin{bmatrix} 1 \\ 0 \end{bmatrix}^{(\tau)}.
\end{equation}

Hence, this verifies the claim made in the main text: $\psitr = \psirt$, showing the top-edge and right-edge Hamiltonians share the same corner mode, as expected for a 2D HOT phase.

\subsection{Model II: \modII~model}
Owing to its similarity with the \modI\  model, the \modII\ model in the presence of a \emph{staggered} Zeeman field $\hs$ can be used to demonstrate the existence of corner states shared between the edge theories. We now elaborate on the discussion from the main text.

\subsubsection{Effective edge theories} \label{subsec:edgehamiltonians-modelII-effedgetheories}

For concreteness, we assume $\K_2 > 0$ and $D > 0$ and study the gap closing that occurs when $\K_1 = \K_2$ and $\hs = \hcstag$, at momentum $\k_0 = (\pi, \pi/2)$. The linearized Hamiltonian in the neighborhood of this point is
\begin{multline}
    \HLEII = 
    \J\sigma_0\eta_0\tau_3 - \K (-\ci\partial_y) \sigma_0\eta_0 \left(\tau_3 - \ci\tau_2\right) 
    \\
    + \Biggl( \frac{\K_1 - \K_2}{2}\sigma_1 - \frac{\K_2}{2} (-\ci\partial_x) \sigma_2 \Biggr) \eta_0 \left(\tau_3 - \ci\tau_2\right)
    \\
    - \sigma_3\eta_2\left( \hs \tau_0 - D\left(\tau_3 - \ci\tau_2\right) \right) 
    \\
    - \Gamma (-\ci\partial_y) \sigma_3\eta_1 \left(\tau_3 - \ci\tau_2\right).
\end{multline}
This equation, with $\K=0$, was reported in the main text in \eqn{eqn:HLEII}.
As explained in the main text, we can analyze the model in the large-$\J$ limit; this way, mass terms analogous to those of \eqn{eqn:HLE-model-I} are easily found. In this limit, we can obtain an effective (Hermitian) Hamiltonian for the positive-energy states using a first-order Schrieffer-Wolff transformation: the unperturbed Hamiltonian is $\J\sigma_0\eta_0\tau_3$, and the subspace of interest is made up of the four particle-like states of the unperturbed Hamiltonian, with the associated projector $\P = \sigma_0 \eta_0 (\tau_0 + \tau_3)/2$. The first-order effective Hamiltonian in $\mathcal{O}(1/\J)$ becomes asymptotically exact in the large-$\J$ limit, and is simply obtained by the projection the full Hamiltonian $\HLEII$ onto the subspace of interest:
\begin{subequations}
\begin{equation}
    \P \Bigl( \HLEII \Bigr) \P = \HLEIIeff \; \P,
\end{equation}
where
\begin{multline}
    \HLEIIeff =
    \bigl( \J - \K (-\ci\partial_y) \bigr) \sigma_0\eta_0 
    \\
    + \biggl( \frac{\K_1 - \K_2}{2}\sigma_1 - \frac{\K_2}{2} (-\ci\partial_x) \sigma_2 \biggr) \eta_0
    \\
    - \sigma_3 \biggl( (\hs-D)\eta_2 + \Gamma (-\ci\partial_y) \eta_1 \biggr).
\end{multline}
\end{subequations}
In this limit, $h_\mathrm{c} \rightarrow D$, so the mass term is manifest: $m_y = \hs - D$. We also let $m_x = (\K_1 - \K_2)/2$, $v_x = \K_2/2$, and $v_y = \Gamma$, so
\begin{multline}
    \HLEIIeff = \bigl( \J - \K (-\ci\partial_y) \bigr) \sigma_0\eta_0 
    \\
    + \bigl( m_x\sigma_1 - v_x (-\ci\partial_x) \sigma_2 \bigr) \eta_0
    \\
    - \sigma_3 \bigl( m_y\eta_2 + v_y (-\ci\partial_y) \eta_1 \bigr).
\end{multline}
The effect of a nonzero $\K$ is to tilt the dispersion along the $k_y$ direction. The outcome of the present calculation is qualitatively the same as long as the Dirac crossing is not inverted (i.e., as long as $|\K| < |v_y| = |\Gamma|$), so we let $\K = 0$ for simplicity.

We create a domain wall along the top edge ($y=0$ line) using a Jackiw-Rebbi ansatz:
\begin{equation}
    \psi_{q_x}(x,y) = \re^{ - \gammatop \int_0^y \rd \tilde{y} \, m_y(\tilde{y})} \re^{ \ci x q_x} \psitop(q_x).
\end{equation}
Independently of the details of $m_y(y)$, the state $\psi_{q_x}(x,y)$ must be an eigenstate of $\HLEIIeff$, meaning
\begin{equation}
    \HLEIIeff \psi_{q_x}(x,y) = E_{q_x} \psi_{q_x}(x,y).
\end{equation}
Hence, $\psitop(q_x)$ must be part of $\mathrm{ker}\bigl(\sigma_3 (\eta_2 + \ci \gammatop v_y \eta_1)\bigr)$. Solutions only arise if $\gammatop = 1 / v_y$, and are of the form
\begin{equation}
    \psitop(q_x) = \chi_{q_x}^{(\sigma)} \otimes \begin{bmatrix} 0 \\ 1 \end{bmatrix}^{(\eta)},
\end{equation}
where $\chi_{q_x}^{(\sigma)}$ is an arbitrary two-component vector. The effective top-edge Hamiltonian is then
\begin{equation}
    \HLEIIefftop = \J \sigma_0 + m_x \sigma_1 - v_x (-\ci\partial_x) \sigma_2.
\end{equation}

We next create a domain wall along the right edge ($x=0$ line) using the following Jackiw-Rebbi ansatz:
\begin{gather}
    \psi_{q_y}(x,y) = \re^{ - \gammaright \int_0^x \rd \tilde{x} \, m_x(\tilde{x})} \re^{ \ci y q_y} \psiright(q_y),
    \\
    \HLEIIeff \psi_{q_y}(x,y) = E_{q_y} \psi_{q_y}(x,y).
\end{gather}
Nontrivial solutions arise only if $\gammaright = 1 / v_x$, and are such that $\psiright(q_y) \in \mathrm{ker}\bigl((\sigma_1 - \ci \sigma_2)\eta_0\bigr)$, meaning
\begin{equation}
    \psiright(q_y) = \begin{bmatrix} 0 \\ 1 \end{bmatrix}^{(\sigma)} \otimes \chi_{q_y}^{(\eta)}.
\end{equation}
This implies the effective Hamiltonian for the right edge is
\begin{equation}
    \HLEIIeffright = \J \eta_0 + m_y \eta_2 + v_y (-\ci\partial_y) \eta_1.
\end{equation}

\subsubsection{Shared corner modes}
We now verify the existence of shared corner modes by introducing domain walls in the edge Hamiltonians $\HLEIIefftop$ and $\HLEIIeffright$.

To find the corner mode originating from the top-edge Hamiltonian, we use the ansatz
\begin{subequations}
    \begin{gather}
        \psitop(x) = \re^{ - \gammatopright \int_0^x \rd \tilde{x} \, m_x(\tilde{x})} \psitr,
        \\
        \psitr = \chi^{(\sigma)} \otimes 
        \begin{bmatrix} 0 \\ 1 \end{bmatrix}^{(\eta)}.
    \end{gather}
\end{subequations}
Such a solution exists if $\gammatopright = 1 / v_x$ and is part of $\mathrm{ker}(\sigma_1 - \ci \sigma_2)$, leading to the following solution at energy $E = \J$:
\begin{equation}
    \psitr = \begin{bmatrix} 0 \\ 1 \end{bmatrix}^{(\sigma)}
        \otimes
        \begin{bmatrix} 0 \\ 1 \end{bmatrix}^{(\eta)}.
\end{equation}

The corner mode originating from the right-edge Hamiltonian is found using the ansatz
\begin{subequations}
    \begin{gather}
        \psiright(y) = \re^{ - \gammarighttop \int_0^y \rd \tilde{y} \, m_y(\tilde{y})} \psirt,
        \\
        \psirt = 
        \begin{bmatrix} 0 \\ 1 \end{bmatrix}^{(\sigma)}
        \otimes
        \chi^{(\eta)}.
    \end{gather}
\end{subequations}
A solution arises if $\gammarighttop = 1 / v_y$ and is part of $\mathrm{ker}(\eta_2 + \ci\eta_1)$. Specifically, 
\begin{equation}
    \psirt = \begin{bmatrix} 0 \\ 1 \end{bmatrix}^{(\sigma)}
        \otimes
        \begin{bmatrix} 0 \\ 1 \end{bmatrix}^{(\eta)},
\end{equation}
at energy $E = \J$.

Comparing $\psitr$ with $\psirt$ we find that they are \emph{identical}, thus confirming that the corner mode is shared between the top and right edges.
     
\section{Invariance under mirror-like symmetries}\label{app:symmetries}
In this section, we obtain the symmetry-allowed spin interactions for the two models above. 
We approach this by considering the symmetry-allowed forms of the tensors $\mathbb{J}^{\br,\br'}$ for all possible bilinear spin interactions $\mathbf{S}_{\br}^\transpose \mathbb{J}^{\br,\br'} \mathbf{S}_{\br'}$ between sites $\br$ and $\br'$. In both cases, we assume the only interaction between bottom and top layers is the strong antiferromagnetic exchange $\J$ within each dimer. Hence, we may drop the layer index and treat each layer individually for the symmetry analysis.

\subsection{Model I: \modI~model}\label{subsec:symmetries_modI}
A priori, there are eight bilinear spin interactions to consider, four of which are intra-unit cell and four, inter-unit cell.
We write the transformations in terms of the matrices $R_x = \mathrm{diag}(1,-1,-1)$ and $R_y = \mathrm{diag}(-1,1,-1)$. 
For notational brevity, we define $\tilde{M}_x := \tilde{C}^z_2\circ M_x$. 
We discuss the symmetry transformations of the interaction tensors on the horizontal and vertical bonds below.

\subsubsection{Intracell bonds}

\paragraph*{Horizontal bonds}
The intracell interaction tensors $\Jintra^{3,1}$ (connecting dimers 3 and 1 in \fig{fig:models}(a) and (b)) and $\Jintra^{2,4}$ (connecting dimers 2 and 4) transform as follows under the symmetries:
\begin{subequations}
\begin{gather}
\begin{split}
    \mathbf{S}_{\R,3}^\transpose \Jintra^{3,1} \mathbf{S}_{\R,1} 
    &\overset{\tilde{M}_x}{\rightarrow}
    \mathbf{S}_{\R',1}^\transpose R_y^\transpose \Jintra^{3,1} R_y \mathbf{S}_{\R',3}
    \\
    &\ \,= \mathbf{S}_{\R',3}^\transpose R_y^\transpose {\Jintra^{3,1}}^\transpose R_y \mathbf{S}_{\R',1};
\end{split}
\\
\begin{split}
    \mathbf{S}_{\R,2}^\transpose \Jintra^{2,4} \mathbf{S}_{\R,4} 
    &\overset{\tilde{M}_x}{\rightarrow}
    \mathbf{S}_{\R',4}^\transpose R_x^\transpose \Jintra^{2,4} R_x \mathbf{S}_{\R',2}
    \\
    &\ \,= \mathbf{S}_{\R',2}^\transpose R_x^\transpose {\Jintra^{2,4}}^\transpose R_x \mathbf{S}_{\R',4};
\end{split}
\\
\begin{split}
    &\mathbf{S}_{\R,3}^\transpose \Jintra^{3,1} \mathbf{S}_{\R,1} + \mathbf{S}_{\R,2}^\transpose \Jintra^{2,4} \mathbf{S}_{\R,4}
    \\
    & \overset{M_y}{\rightarrow}
    \mathbf{S}_{\R'',2}^\transpose R_y^\transpose \Jintra^{3,1} R_y \mathbf{S}_{\R'',4} + \mathbf{S}_{\R'',3}^\transpose  R_y^\transpose  \Jintra^{2,4} R_y \mathbf{S}_{\R'',1}.
\end{split}
\end{gather}
\end{subequations}
Hence, the symmetries constrain these exchange tensors as follows:
\begin{subequations}
\begin{gather}
    \Jintra^{3,1} \overset{\tilde{M}_x}{=} R_y^\transpose {\Jintra^{3,1}}^\transpose R_y;
    \\
    \Jintra^{2,4} \overset{\tilde{M}_x}{=} R_x^\transpose {\Jintra^{2,4}}^\transpose R_x;
    \\
    \Jintra^{3,1} \overset{M_y}{=} R_y^\transpose \Jintra^{2,4} R_y;
    \ \ 
    \Jintra^{2,4} \overset{M_y}{=} R_y^\transpose \Jintra^{3,1} R_y.
\end{gather}
\end{subequations}
This leads to the following explicit structure for the tensors:
\begin{align}\label{eq:symallowed_intra_horiz}
    \Jintra^{3,1} \!\!=\!\! \begin{bmatrix}
        \hphantom{-} a_{11} & a_{12} & \\
        -a_{12} & a_{22} & \\
        & & a_{33}
    \end{bmatrix}\!\!,\ 
    \Jintra^{2,4} \!\!=\!\! \begin{bmatrix}
        a_{11} & -a_{12} & \\
        a_{12} & \hphantom{-}a_{22} & \\
        & & a_{33}
    \end{bmatrix}\!\!.
\end{align}
In our model, we have $a_{12} = 0$ and $a_{11} = a_{22} = a_{33} = \Kh_1$, which is compatible with the symmetry constraints.

\paragraph*{Vertical bonds}
The intracell interaction tensors $\Jintra^{2,3}$ (connecting dimers 2 and 3 in \fig{fig:models}(b)) and $\Jintra^{4,1}$ (connecting dimers 4 and 1) transform as follows:
\begin{subequations}
\begin{gather}
\begin{split}
    \mathbf{S}_{\R,2}^\transpose \Jintra^{2,3} \mathbf{S}_{\R,3} 
    &\overset{M_y}{\rightarrow}
    \mathbf{S}_{\R',3}^\transpose R_y^\transpose \Jintra^{2,3} R_y \mathbf{S}_{\R',2}
    \\
    &\ \,= \mathbf{S}_{\R',2}^\transpose R_y^\transpose {\Jintra^{2,3}}^\transpose R_y \mathbf{S}_{\R',3};
\end{split}
\\
\begin{split}
    \mathbf{S}_{\R,4}^\transpose \Jintra^{4,1} \mathbf{S}_{\R,1} 
    &\overset{M_y}{\rightarrow}
    \mathbf{S}_{\R',1}^\transpose R_y^\transpose \Jintra^{4,1} R_y \mathbf{S}_{\R',4}
    \\
    &\ \,= \mathbf{S}_{\R',4}^\transpose R_y^\transpose {\Jintra^{4,1}}^\transpose R_y \mathbf{S}_{\R',1};
\end{split}
\\
\begin{split}
    &\mathbf{S}_{\R,2}^\transpose \Jintra^{2,3} \mathbf{S}_{\R,3} + \mathbf{S}_{\R,4}^\transpose \Jintra^{4,1} \mathbf{S}_{\R,1}
    \\
    &\overset{\tilde{M}_x}{\rightarrow}
    \mathbf{S}_{\R'',4}^\transpose R_x^\transpose \Jintra^{2,3} R_y \mathbf{S}_{\R'',1} + \mathbf{S}_{\R'',2}^\transpose R_x^\transpose \Jintra^{4,1} R_y \mathbf{S}_{\R'',3}.
\end{split}
\end{gather}
\end{subequations}
Hence, the symmetry constraints on these tensors are
\begin{subequations}
\begin{gather}
    \Jintra^{2,3} \overset{M_y}{=} R_y^\transpose {\Jintra^{2,3}}^\transpose R_y;
    \\
    \Jintra^{4,1} \overset{M_y}{=} R_y^\transpose {\Jintra^{4,1}}^\transpose R_y;
    \\
    \Jintra^{2,3} \overset{\tilde{M}_x}{=} R_x^\transpose \Jintra^{4,1} R_y;
    \ \ 
    \Jintra^{4,1} \overset{\tilde{M}_x}{=} R_x^\transpose \Jintra^{2,3} R_y,
\end{gather}
\end{subequations}
and their explicit structure is
\begin{align}\label{eq:symallowed_intra_verti}
    \Jintra^{2,3} \!\!=\!\! \begin{bmatrix}
        -b_{11} & \hphantom{-}b_{12} & \\
        -b_{12} & -b_{22} & \\
        & & b_{33}
    \end{bmatrix}\!\!,\ 
    \Jintra^{4,1} \!\!=\!\! \begin{bmatrix}
        \hphantom{-}b_{11} & b_{12} & \\
        -b_{12} & b_{22} & \\
        & & b_{33}
    \end{bmatrix}\!\!.
\end{align}
In our model, we have $b_{12} = 0$ and $b_{11} = b_{22} = b_{33} = \Kv_1$, making it compatible with the symmetry constraints.

\subsubsection{Intercell bonds}
The symmetry constraints for the tensors on the intercell bonds can also be derived using the previous analysis.
At the end of the analysis, we find that the horizontal intercell tensors $\Jinter^{1,3}$ and $\Jinter^{4,2}$ have the same form (\eqn{eq:symallowed_intra_horiz}) as the horizontal intracell tensors $\Jintra^{3,1}$ and $\Jintra^{2,4}$, respectively. Likewise, the vertical intercell tensors $\Jinter^{1,4}$ and $\Jinter^{3,2}$ have the same structure (\eqn{eq:symallowed_intra_verti}) as the tensors $\Jintra^{4,1}$ and $\Jintra^{2,3}$, respectively.
Specifically, 
\begin{align}\label{eq:symallowed_inter_horiz}
    \Jinter^{1,3} &= \begin{bmatrix}
        \hphantom{-} c_{11} & \hphantom{-} c_{12} & \\
        -c_{12} & \hphantom{-} c_{22} & \\
        & & c_{33}
    \end{bmatrix},
    &
    \Jinter^{4,2} &= \begin{bmatrix}
        c_{11} & -c_{12} & \\
        c_{12} & \hphantom{-} c_{22} & \\
        & & c_{33}
    \end{bmatrix},
    \\
    \Jinter^{3,2} &= \begin{bmatrix}
        -d_{11} & \hphantom{-} d_{12} & \\
        -d_{12} & -d_{22} & \\
        & & d_{33}
    \end{bmatrix},
    &
    \Jinter^{1,4} &= \begin{bmatrix}
        \hphantom{-} d_{11} & d_{12} & \\
        -d_{12} & d_{22} & \\
        & & d_{33}
    \end{bmatrix}.
\end{align}
For the present model, we have $c_{12} = 0$, $c_{11} = c_{22} = c_{33} = \Kh_2$, $d_{12} = 0$, and $d_{11} = d_{22} = d_{33} = \Kv_2$, making the model compatible with the symmetry constraints for the intercell bonds as well.

\subsection{Model II: \modII~model}\label{subsec:symmetries_modII}
In the \modII\ model, we consider four bilinear spin interactions, two of which are horizontal and two, vertical. To describe the transformations, we define the matrices
\begin{align}
    R_x &= \begin{bmatrix}
    1 & & \\
    & -1 & \\
    & & -1
    \end{bmatrix},
    &
    \tilde{R}_y &= \begin{bmatrix}
    0 & 1 & \\
    1 & 0 & \\
    & & -1
    \end{bmatrix}.
\end{align}

\subsubsection{Horizontal bonds}
The interaction tensors $\KI$ (between $\S_{\R,1}$ and $\S_{\R,2}$; see \fig{fig:models}(c) and (d)) and $\KII$ (between $\S_{\R,2}$ and $\S_{\R+\hat{x},1}$) transform as follows under the symmetries:
\begin{subequations}
\begin{gather}
\begin{split}
    \mathbf{S}_{\R,1}^\transpose \KI \mathbf{S}_{\R,2}
    & \overset{M_x}{\rightarrow}
    \mathbf{S}_{\R',2}^\transpose R_x^\transpose \KI R_x \mathbf{S}_{\R',1}
    \\
    &\ \, = \mathbf{S}_{\R',1}^\transpose R_x^\transpose \KI^\transpose R_x \mathbf{S}_{\R',2},
\end{split}
\\
\begin{split}
    \mathbf{S}_{\R,1}^\transpose \KI \mathbf{S}_{\R,2}
    & \overset{\tilde{M}_y}{\rightarrow}
    \mathbf{S}_{\R'',1}^\transpose \tilde{R}_y^\transpose \KI \tilde{R}_y \mathbf{S}_{\R'',2};
\end{split}
\\
\begin{split}
    \mathbf{S}_{\R,2}^\transpose \KII \mathbf{S}_{\R+\hat{x},1}
    & \overset{M_x}{\rightarrow}
    \mathbf{S}_{\R',1}^\transpose R_x^\transpose \KII \mathbf{S}_{\R'-\hat{x},2}
    \\
    &\ \, = \mathbf{S}_{\R',2}^\transpose R_x^\transpose {\KII}^\transpose \mathbf{S}_{\R'+\hat{x},1},
\end{split}
\\
\begin{split}
    \mathbf{S}_{\R,2}^\transpose \KII \mathbf{S}_{\R+\hat{x},1}
    & \overset{\tilde{M}_y}{\rightarrow}
    \mathbf{S}_{\R'',2}^\transpose \tilde{R}_y^\transpose \KII \tilde{R}_y \mathbf{S}_{\R''+\hat{x},1},
\end{split}
\end{gather}
\end{subequations}
constraining these exchange tensors as follows:
\begin{subequations}
\begin{align}
    \KI &\overset{M_x}{=} R_x^\transpose \KI^\transpose R_x,
    &
    \KI &\overset{\tilde{M}_y}{=} \tilde{R}_y^\transpose \KI \tilde{R}_y;
    \\
    \KII &\overset{M_x}{=} R_x^\transpose \KII^\transpose R_x,
    &
    \KII &\overset{\tilde{M}_y}{=} \tilde{R}_y^\transpose \KII \tilde{R}_y.
\end{align}
\end{subequations}
The explicit form of the interaction tensors is
\begin{align}
    \KI &= \mathrm{diag} ( a_{11}, a_{11}, a_{33} ),
    &
    \KII &= \mathrm{diag} ( b_{11}, b_{11}, b_{33} ).
\end{align}
In our model, $a_{11} = a_{33} = \K_1$ and $b_{11} = b_{33} = \K_2$, making it exactly compatible with the symmetry conditions.

\subsubsection{Vertical bonds}
The interaction terms $\C$ (between $\S_{\R,1}$ and $\S_{\R+\hat{y},1}$; see \fig{fig:models}(c) and (d)) and $\D$ (between $\S_{\R,2}$ and $\S_{\R+\hat{y},2}$) transform as follows under the symmetries.
\begin{subequations}
\begin{gather}
\begin{split}
    \mathbf{S}_{\R,1}^\transpose \C \mathbf{S}_{\R+\hat{y},1}
    & \overset{\tilde{M}_y}{\rightarrow}
    \mathbf{S}_{\R'',1}^\transpose \tilde{R}_y^\transpose \C \tilde{R}_y \mathbf{S}_{\R''-\hat{y},1}
    \\
    &=
    \mathbf{S}_{\R''-\hat{y},1}^\transpose \tilde{R}_y^\transpose \C^\transpose \tilde{R}_y \mathbf{S}_{\R'',1}
    \\
    &=
    \mathbf{S}_{\R'',1}^\transpose \tilde{R}_y^\transpose \C^\transpose \tilde{R}_y \mathbf{S}_{\R''+\hat{y},1};
\end{split}
\\
\begin{split}
    \mathbf{S}_{\R,2}^\transpose \D \mathbf{S}_{\R+\hat{y},2}
    & \overset{\tilde{M}_y}{\rightarrow}
    \mathbf{S}_{\R'',2}^\transpose \tilde{R}_y^\transpose \D \tilde{R}_y \mathbf{S}_{\R''-\hat{y},2}
    \\
    &=
    \mathbf{S}_{\R''-\hat{y},2}^\transpose \tilde{R}_y^\transpose \D^\transpose \tilde{R}_y \mathbf{S}_{\R'',2}
    \\
    &=
    \mathbf{S}_{\R'',2}^\transpose \tilde{R}_y^\transpose \D^\transpose \tilde{R}_y \mathbf{S}_{\R''+\hat{y},2};
\end{split}
\\
\begin{split}
    &\mathbf{S}_{\R,1}^\transpose \C \mathbf{S}_{\R+\hat{y},1} + \mathbf{S}_{\R,2}^\transpose \D \mathbf{S}_{\R+\hat{y},2}
    \\
    &\overset{M_x}{\rightarrow} \!
    \mathbf{S}_{\R',2}^\transpose \! R_x^\transpose \! \C R_x \mathbf{S}_{\R'+\hat{y},2} \!+\! \mathbf{S}_{\R',1}^\transpose \! R_x^\transpose \! \D R_x \mathbf{S}_{\R'+\hat{y},1}.
\end{split}
\end{gather}
\end{subequations}
The exchange tensors must obey the following symmetry constraints.
\begin{subequations}
\begin{gather}
\begin{split}
    \C \overset{M_x}{=} R_x^\transpose \D R_x,;
    \ \ 
    \D \overset{M_x}{=} R_x^\transpose \C R_x;
\end{split}
\\
\begin{split}
    \C \overset{\tilde{M}_y}{=} \tilde{R}_y^\transpose \C^\transpose \tilde{R}_y;
\end{split}
\\
\begin{split}
    \D \overset{\tilde{M}_y}{=} \tilde{R}_y^\transpose \D^\transpose \tilde{R}_y.
\end{split}
\end{gather}
\end{subequations}
These constraints yield the following explicit form for the interaction tensors.
\begin{align}
    \C &= \begin{bmatrix}
        \hphantom{-} c_{11} & -c_{12}& \\ -c_{21} & \hphantom{-} c_{11}& \\ &&c_{33}
    \end{bmatrix},
    &
    \D &= \begin{bmatrix}
        c_{11}&c_{12}& \\ c_{21}&c_{11}& \\ &&c_{33}
    \end{bmatrix}
\end{align}
In our model, $c_{11} = c_{33} = \K$, $c_{12} = \Gamma + D$ and $c_{21} = \Gamma - D$, making it exactly compatible with the symmetries.

\subsubsection{Zeeman field}

As stated in the main text, the symmetries $M_x$ and $\tilde{M}_y$ do not allow a Zeeman field. However, as we shall see, certain Zeeman fields are allowed if one or both symmetry operations are modified to include time reversal. Note that the other interactions in the spin Hamiltonian are invariant under time reversal because they are all bilinear in spins with real coefficients. 

We consider a general (translation-invariant) Zeeman coupling term, $\sum_{a\in\{1,2\}} \mathbf{B}_a^\transpose \mathbf{S}_{\R,a}$, where $\mathbf{B}_a = [B_a^x, B_a^y, B_a^z]^\transpose$. 
Under the operation $\tilde{M}_y$, the Zeeman term transforms as
\begin{equation}
    \mathbf{B}_a^\transpose \mathbf{S}_{\R,a}
    \overset{\tilde{M}_y}{\rightarrow}
    \mathbf{B}_a^\transpose \tilde{R}_y \mathbf{S}_{\R'',a},
\end{equation}
leading to the constraint $\mathbf{B}_a \overset{\tilde{M}_y}{=} \tilde{R}_y^\transpose \mathbf{B}_a$. This constraint implies that $\mathbf{B}_a$ must be an eigenvector of $\tilde{R}_y^\transpose$ with eigenvalue 1. The only such eigenvector is $[1, 1, 0]^\transpose$, meaning $B_a^x = B_a^y$ and $B_a^z = 0$ for $a \in \{1,2\}$.

On the other hand, under $M_x$, the transformation proceeds as
\begin{equation}
    \mathbf{B}_1^\transpose \mathbf{S}_{\R,1} \!+\! \mathbf{B}_2^\transpose \mathbf{S}_{\R,2}
    \overset{M_x}{\rightarrow}
    \mathbf{B}_1^\transpose R_x \mathbf{S}_{\R',2} \!+\! \mathbf{B}_2^\transpose R_x \mathbf{S}_{\R',1},
\end{equation}
giving rise to the constraint $\mathbf{B}_1 \overset{M_x}{=} R_x^\transpose \mathbf{B}_2$, $\mathbf{B}_2 \overset{M_x}{=} R_x^\transpose \mathbf{B}_1$. 
Explicitly, this implies $( B_1^x, B_1^y, B_1^z ) = ( B_2^x, -B_2^y, -B_2^z )$.

Under the modified operation $\TR\circ\tilde{M}_y$, the Zeeman terms transform as
\begin{equation}
    \mathbf{B}_a^\transpose \mathbf{S}_{\R,a}
    \overset{\TR\circ\tilde{M}_y}{\rightarrow}
    - \mathbf{B}_a^\transpose \tilde{R}_y \mathbf{S}_{\R'',a},
\end{equation}
leading to the constraint $\mathbf{B}_a \overset{\TR\circ\tilde{M}_y}{=} -\tilde{R}_y^\transpose \mathbf{B}_a$. 
Since the eigenvectors of $\tilde{R}_y^\transpose$ with eigenvalue $-1$ are $[1, -1, 0]^\transpose$ and $[0, 0, 1]^\transpose$, this constraint forces $B_a^x = - B_a^y$ and leaves $B_a^z$ unconstrained, where $a \in \{1,2\}$.

Finally, the spins transform as follows under $\TR \circ M_x$:
\begin{equation}
    \mathbf{B}_1^\transpose \mathbf{S}_{\R,1} + \mathbf{B}_2^\transpose \mathbf{S}_{\R,2}
    \overset{\TR \circ M_x}{\rightarrow}
    - \mathbf{B}_1^\transpose R_x \mathbf{S}_{\R',2} - \mathbf{B}_2^\transpose R_x \mathbf{S}_{\R',1},
\end{equation}
This gives rise to the constraint $\mathbf{B}_1 \overset{M_x}{=} - R_x^\transpose \mathbf{B}_2$, $\mathbf{B}_2 \overset{M_x}{=} - R_x^\transpose \mathbf{B}_1$, and implies $( B_1^x, B_1^y, B_1^z ) = ( - B_2^x, B_2^y, B_2^z )$

The symmetry-allowed Zeeman terms for the different combination of symmetries are summarized in Table~\ref{tab:B_constraints}, and verify the assertions made in the main text.
\begin{table}
\begin{tabular}{c||c|c}
  & $\tilde{M}_y$ & $\TR \circ \tilde{M}_y$ 
  \\ \hline \hline
      $M_x$
      & 
      $\mathbf{B}_1 = \mathbf{B}_2 = 0$
      &
      $\mathbf{B}_1 = - \mathbf{B}_2 = \hs \hat{z}$
  \\ \hline
      $\TR \circ M_x$
      & 
      $\mathbf{B}_1 = \mathbf{B}_2 = 0$
      &
      $\mathbf{B}_1 = + \mathbf{B}_2 = h\, \hat{z}$
\end{tabular}
\caption{Symmetry constraints on Zeeman field.}
\label{tab:B_constraints}
\end{table}

\subsection{Implementation on triplons}
Any symmetry of the spin Hamiltonian that maps one dimer to another has a representation in the triplon Hamiltonian.
\paragraph{Unitary transformations}
If a symmetry $g$ transforms spins such that
\begin{equation}
    S_{i,\ell}^\alpha \overset{g}{\rightarrow} \sum_\beta T_{\alpha\beta} S_{i',\ell'}^\beta
\end{equation}
for some SO(3) matrix $T$, its transformation on the triplon operators is
\begin{equation}
    t_{i,\alpha} \rightarrow 
    \begin{cases}
        \hphantom{-} \sum_\beta T_{\alpha\beta} t_{i',\beta} &\mbox{if } \ell=\ell'
        \\
        -\sum_\beta T_{\alpha\beta} t_{i',\beta} &\mbox{if } \ell \neq \ell'
    \end{cases}.
\end{equation}
For the symmetries discussed in \Sec{subsec:symmetries_modI} and \ref{subsec:symmetries_modII}, the $\ell = \ell'$ case in the above equation applies.
Using this prescription and the transformations for the spins, it is straightforward to find the representations given in \eqns{eq:m_matrices_model1} and \ref{eqn:symm-rep-trip-II}.

\paragraph{Antiunitary transformations}

Given the definition of \eqn{eq:spin_triplon_decomposition}, the real-space triplon operators $t^\pdg_{\br \alpha}$ are invariant under time reversal $\TR$; hence, the action of time reversal on a real-space triplon Hamiltonian (e.g.\ \eqns{eqn:trip-ham-real-I} and \ref{eqn:trip-ham-real-II}) is simply to complex-conjugate the coefficients~\mycite{penc2011PRB}. 
From this, we see that the implementation of an antiunitary symmetry transformation $\TR \circ g$, where $g$ is a unitary transformation, is the same as that of $g$ but with an additional (left-multiplied) complex conjugation $\cconj$. For example, the transformations $\TR \circ M_x$ and $\TR \circ \tilde{M}_y = \TR \circ  C^z_4 \circ M_y$ of the \modII\ model have the following implementations:
\begin{subequations}\label{eqn:symm-rep-trip-II-TR}
\begin{align}
    \TR \circ M_x &\equiv \cconj \sigma_1 \eta_3 \tau_0,
    \\
    \TR \circ \tilde{M}_y &\equiv \cconj \sigma_0 \eta_1 \tau_0,
\end{align}
\end{subequations}
c.f.\ \eqn{eqn:symm-rep-trip-II}.

Apart from Zeeman terms, the coefficients in real-space triplon Hamiltonians (e.g.\ \eqns{eqn:trip-ham-real-I} and \ref{eqn:trip-ham-real-II}) are purely real, 
so the representation of $\TR \circ g$, with its additional complex conjugation $\cconj$, has precisely the same effect as the representation of $g$. This is a manifestation of the fact that symmetries symmetries $g$ and $\TR \circ g$ are equivalent for bilinear spin interactions.

In the analysis of the \modII\ model above, we found that the staggered Zeeman field configuration is compatible with symmetries $M_x$ and $\TR \circ \tilde{M}_y$, while the uniform field configuration is compatible with $\TR \circ M_x$ and $\TR \circ \tilde{M}_y$. In the associated real-space triplon Hamiltonian, the Zeeman coupling enters in the intra-unit cell coupling matrix $M_0$ of \eqn{eqn:trip-ham-real-II} in the form of an additional term $-\hs \sigma_3 \eta_2 \tau_3$ (staggered field) or $-h \sigma_0 \eta_2 \tau_3$ (uniform field). 
Using to the representations of \eqns{eqn:symm-rep-trip-II} and \ref{eqn:symm-rep-trip-II-TR}, it is easy to see that the Zeeman terms remain invariant in the following way.
For the staggered field and the symmetries $M_x$ and $\TR \circ \tilde{M}_y$, we have
\begin{equation}
\begin{split}
    (\sigma_1 \eta_3 \tau_0)^{-1} (\!-\hs \sigma_3 \eta_2 \tau_3) (\sigma_1 \eta_3 \tau_0) &\!=\! -\hs \sigma_3 \eta_2 \tau_3
    \\
    (\!\cconj \sigma_0 \eta_1 \tau_0)^{-1} (\!-\hs \sigma_3 \eta_2 \tau_3) (\!\cconj \sigma_0 \eta_1 \tau_0) &\!=\! -\hs \sigma_3 \eta_2 \tau_3,
\end{split}
\end{equation}
while for the uniform field and the symmetries $\TR \circ M_x$ and $\TR \circ \tilde{M}_y$,
\begin{equation}
\begin{split}
    (\cconj \sigma_1 \eta_3 \tau_0)^{-1} (-h \sigma_0 \eta_2 \tau_3) (\cconj \sigma_1 \eta_3 \tau_0) &= -h \sigma_0 \eta_2 \tau_3
    \\
    (\cconj \sigma_0 \eta_1 \tau_0)^{-1} (-h \sigma_0 \eta_2 \tau_3) (\cconj \sigma_0 \eta_1 \tau_0) &= -h \sigma_0 \eta_2 \tau_3.
\end{split}
\end{equation}
Since the remaining terms in the real-space triplon Hamiltonian (resulting from the bilinear spin intractions) remain invariant by the previously stated argument, the overall real-space triplon Hamiltonians with staggered and uniform Zeeman fields are invariant with respect to $M_x$, $\TR \circ \tilde{M}_y$ and $\TR \circ M_x$, $\TR \circ \tilde{M}_y$, respectively.

Finally, as claimed in the main text, we see that the representations of the symmetries $M_x$ and $\TR \circ \tilde{M}_y$ appropriate to the staggered field do indeed anticommute:
\begin{equation}
    \big\{
    \sigma_1 \eta_3 \tau_0, 
    \cconj \sigma_0 \eta_1 \tau_0
    \big\}
    = 0.
\end{equation}
Using this prescription and the transformations for the spins, it is straightforward to find the representations given in \eqns{eq:m_matrices_model1} and \ref{eqn:symm-rep-trip-II}.

\section{Effect of exchange disorder on the corner modes}

We have investigated the robustness of the corner modes against various types of disorder in finite-size samples. 
We find that weak disorder that preserves the symmetries on average do not destroy the corner modes. 
Specifically, we introduce disorder of strength $w$ by uniformly sampling the Heisenberg interactions on each interdimer bond from the interval $[J-w/2,J+w/2]$, where $J\in \{\Kh_{1,2}, \Kv_{1,2}\}$.
Indeed, \fig{fig:disorder} illustrates that increasing disorder in the \modI\ model increases the penetration depth of the corner states into the bulk (see first and second rows of \fig{fig:disorder}), but only when the disorder is strong enough to close the bandgap do the corner modes become delocalized (see third row of \fig{fig:disorder}).
A similar analysis on the \modII\ model yielded analogous results.

\begin{figure}
    \centering
\includegraphics[scale=0.99]{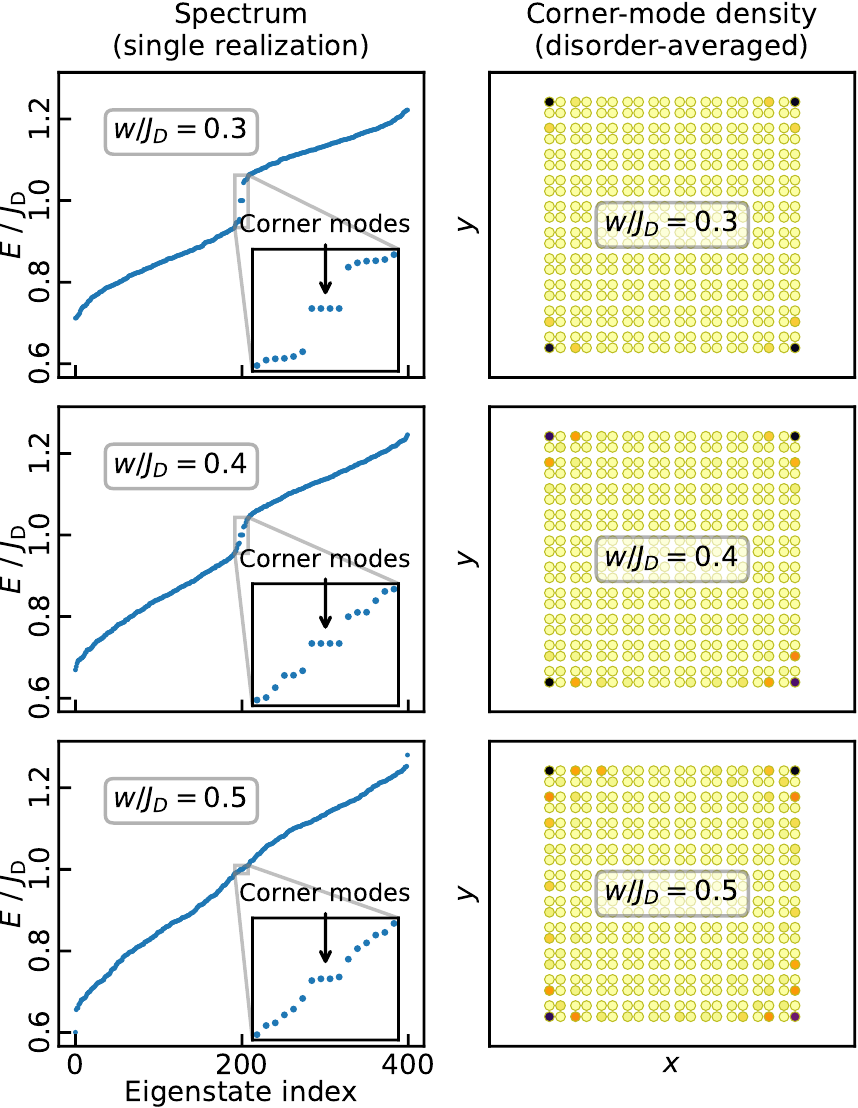}
\caption{Spectrum for a single disorder realization (left) and disorder-averaged total corner-mode density (right) for disorder of the form described in the main text, and for three different disorder strengths ($w/\J=0.3$, $w/\J=0.4$, $w/\J=0.5$). The corner modes become delocalized once the disorder is strong enough to close the bandgap. The mean parameter values for the inter-dimer Heisenberg-interactions are $\overline{\Kh_1}=\overline{\Kv_1}=0.02 \J, \overline{\Kh_2}=\overline{\Kv_2}=0.2 \J$.}
  \label{fig:disorder}
\end{figure}
 \fi

\fi
\bibliographystyle{apsrev4-2}
\bibliography{mainbib}

\end{document}